\documentstyle[aas2pp4,twoside]{article}

\newcommand{\uK}{\,\mbox{$\mu{\rm K}\ $}}
\newcommand{\PuK}{\phantom{\uK}} 
\newcommand{\Ka}{K$_{\rm a}$}
\newcommand{\B}{\phantom{0}}
\newcommand{\AB}{\mbox{$A-B$}}
\newcommand{\EW}{\mbox{$E-W$}}
\newcommand{\HH}{\mbox{$H1-H2$}}
\newcommand{\MSOL}{SK95/MSAM Overlap}
\newcommand{\MS}{MSAM1-92}
\newcommand{\RING} {\it Ring \rm}
\newcommand{\CAP} {\it Cap \rm}

\slugcomment{PRELIMINARY}

\lefthead{Netterfield et al.} 
\righthead{The Angular Power Spectrum of the CMB}

\begin{document}

\title{A Measurement of the Angular Power Spectrum of the Anisotropy in
  the Cosmic Microwave Background}

\author{C. B. Netterfield, M. J. Devlin, N. Jarosik, L. Page 
  and E. J. Wollack\altaffilmark{1}}
\affil{Princeton University, Department of Physics, Jadwin Hall, \\
  Princeton, NJ, 08544}


\altaffiltext{1}{NRAO, 2015 Ivy Rd., Charlottesville, VA, 22903}


\begin{abstract}
  We report on a measurement of the angular power spectrum of the
  anisotropy in the Cosmic Microwave Background.  The anisotropy is
  measured in 23 different multipole bands from $\ell=54$ ($\approx
  3\arcdeg$) to $\ell=404$ ($\approx 0.45\arcdeg$) and in 6 frequency
  bands from 26\,GHz to 46\,GHz over three observing seasons.  The
  measurements are consistent from year to year.  The frequency spectral
  index of the fluctuations (measured at low $\ell$) is consistent with
  that of the CMB and inconsistent with either dust or Galactic
  free-free emission.  Furthermore, the observations of the \MS\ 
  experiment (\cite{che94}) are repeated and confirmed.  The angular
  spectrum shows a distinct rise from $\delta T_\ell \equiv
  \sqrt{\ell(2\ell+1)<|a^m_\ell|^2>/4\pi} = 49^{+8}_{-5}$\uK at
  $\ell=87$ to $\delta T_\ell=85^{+10}_{-8}\uK$ at $\ell = 237$.  These
  values do not include an overall $\pm 14\%$ $(1\sigma)$ calibration
  uncertainty.  The analysis and possible systematic errors are
  discussed.
\end{abstract}




%

 \section{Introduction}
 \pagestyle{myheadings}
 The discovery of the anisotropy in the cosmic microwave background
 (CMB) by the COBE satellite (\cite{smo92}; \cite{Ben92}) and its
 subsequent confirmation (\cite{gan93}) at large angular scales (greater
 than $7\arcdeg$) has been followed by positive detections of
 anisotropies in the microwave sky at intermediate angular scales by a
 number of other experiments (\cite{DeB94}; \cite{Dev94}; \cite{Dra94};
 \cite{Gun95}; \cite{Gut95}) and upper limits at smaller angular scales
 (\cite{Mey93}; \cite{Tuc93}). In order to reject foreground
 contamination, most of these experiments observe the sky over a range
 of frequencies.  Additionally, several have successfully repeated their
 measurements of previous years.

 Characterization of the anisotropy at medium angular scales ($2\arcdeg$
 to $0.2\arcdeg$) can strongly constrain theories of structure formation
 and cosmological parameters, mainly through the angular power spectrum
 of the anisotropy (\cite{Bon94b}; \cite{Cri95}; \cite{Kam94};
 \cite{Jun95}.  For a recent review of the CMB see \cite{Whi94b}).

 We present results from an experiment designed to measure the angular
 power spectrum of the CBR at medium angular scales.  The SK telescope
 observes from the ground in Saskatoon, Saskatchewan, Canada.  The
 observing scheme has many internal consistency tests which allow checks
 of the integrity of the measurement.  A frequency span of 20\,GHz
 provides discrimination against foreground contaminants.  Measurements
 made by this experiment are consistent from year to year.  In addition,
 we have reproduced the results of the \MS\ (\cite{che94}) experiment.
 Partial descriptions of the SK instrument are given in \cite{Wol93}
 (SK93), \cite{Wol94} (SK93), \cite{Wol94b}, \cite{Pag94}, \cite{Net94}
 (SK94) and \cite{Net95} (SK93 to SK94 comparison).  The instrument and
 calibration is described in detail in \cite{Wol96}.  Previous results
 (SK93 and SK94) are included in this analysis.

 \section{Instrument and Calibration}
 The SK telescope is comprised of a corrugated feed horn illuminating a
 parabolic primary followed by a chopping flat with a vertical chopping
 axis.  The telescope is steerable in azimuth, but fixed in elevation.
 Observations are made in \Ka-band (26 GHz to 36 GHz) and Q-band (36 GHz
 to 46 GHz).  Observations made in each band are broken up into 3
 frequency sub-bands and two linear polarizations.  Total power
 radiometers based on HEMT amplifiers are used.  The supernova remnant
 Cas-A is used to calibrate the telescope and determine the beam size
 and pointing.  Table \ref{table:obParms} lists the beam parameters and
 uncertainties.  The beam widths are known to better than $2\%$.  The
 calibration uncertainty is comprised of a $13\%$ contribution due to
 uncertainty in the absolute calibration of Cas-A, which is constant
 between years and radiometers, and a $3\%$ to $5\%$ contribution due to
 measurement uncertainty.

 The telescope pointing is determined from the position of Cas-A.  In
 1995, the telescope was pointed $0.05\arcdeg$ differently than was
 intended, which was included in the analysis.  In 1994, the discrepancy
 was less than 1/10 of a beam, and was ignored.  In addition, there was
 a $0.03\arcdeg\ (1\sigma)$ jitter in the beam position from the chopping plate
 and base pointing accuracy.  When convolved with the beam, this can be
 approximated by adding this uncertainty in quadrature to the nominal
 beam width.  For 1995 this is results in a $1\%$ widening of the
 effective azimuth beam.

 \section {Observing Strategy: Synthesized Beams} \label{sec:beams}
 
 \begin{figure}[tbhp]
   \plotone{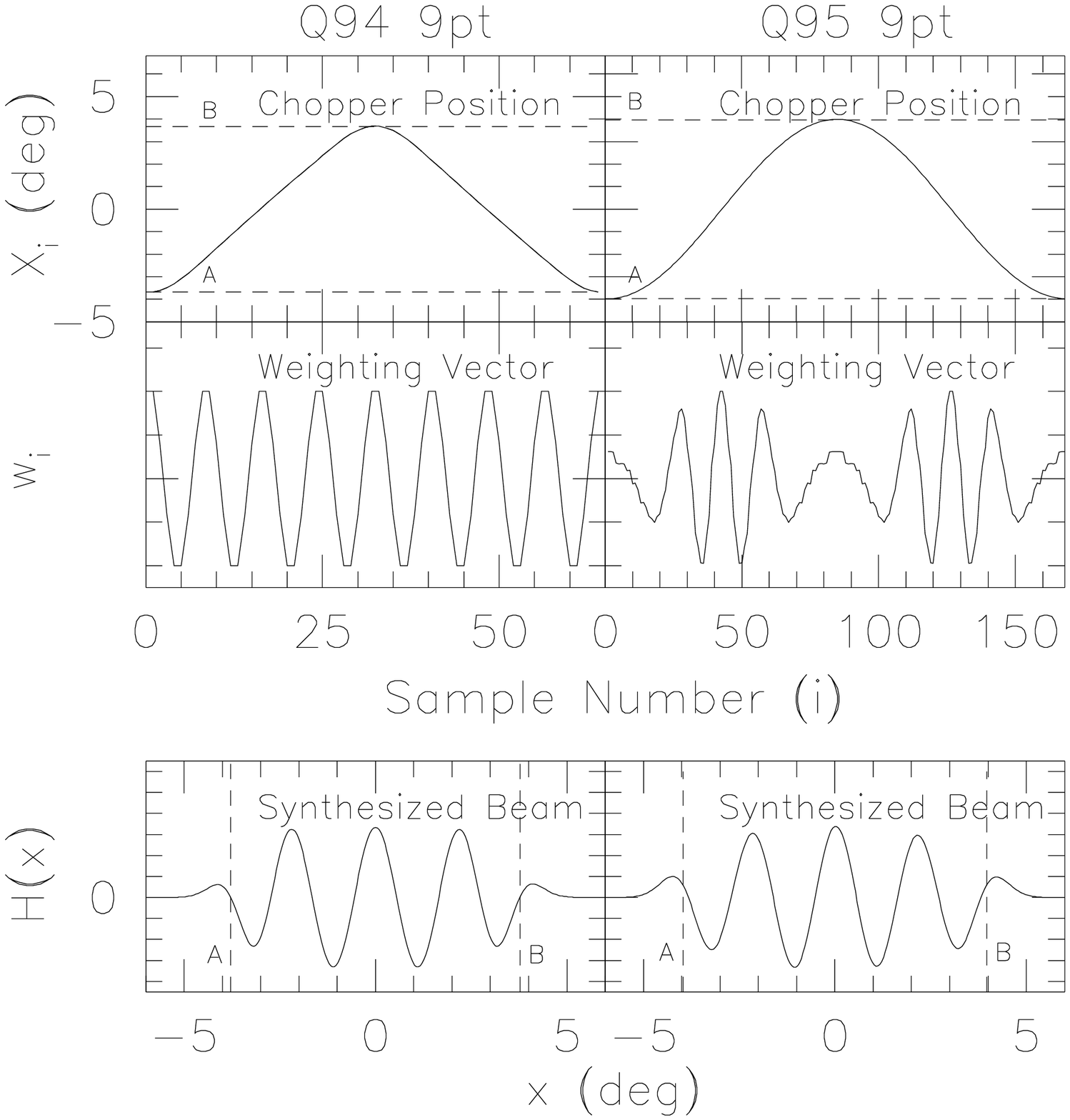}
   \caption{Beam Synthesis.  Shown is the synthesis of the 9pt
     beams for Q94 and Q95 using equation (\ref{eq:antenna}).  $H(x)$ is shown
     along the path of the swept beam.  The effect of binning in RA is
     neglected here for clarity.  In 1994, an approximately linear sweep
     pattern, ${\bf X}_i$, was used as shown.  This allowed the use
     of sinusoidal weighting vectors, $w_i$, to produce synthesized
     beams with equally spaced lobes.  The use of a sinusoidal sweep
     rendered this impossible in 1995.  The Q95 weighting vectors were
     generated by optimizing the sensitivity of the synthesized beam to
     a specified band of angular scales.  Spatially similar synthesized
     beams are produced.}
   \label{fig:synth}
 \end{figure}

 \begin{figure}[tbhp]
   \plotone{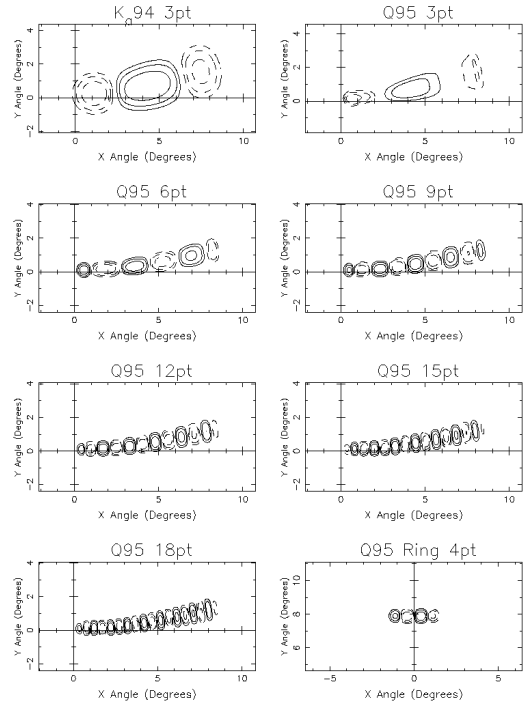}
 \caption{Synthesized Beams.  The contour lines are every 3dB.
   Dashed lines denote negative weighting.  The NCP is located at (0,0).
   Straight lines from the NCP are lines of constant hour angle.  The
   bottom right plot is of the \RING 4pt synthesized beam, which
   is centered $8\arcdeg$ above the NCP.  The remaining plots are of
   \CAP data acquired in the East.  The curvature is due to the
   fact that the beam is swept in azimuth, and not declination.  The
   lobes farthest from the NCP are wider in RA than those closest
   because the data are averaged into 24 (3pt to 5pt) or 48 RA bins.}
\label{fig:antennas}
\end{figure}

\begin{deluxetable}{rlccccc}
\tablecaption{Observing Parameters \label{table:obParms}}
\tablehead{
\colhead{}&\colhead{}& \colhead{\Ka93} & \colhead{\Ka94} &
\colhead{Q94} & \colhead{Q95 Cap} & \colhead{Q95 RING} }
\startdata
Beam    & $x$ FWHM & $1.42 \pm 0.005 \arcdeg$ & $1.42 \pm 0.005 \arcdeg$ &
                $1.004\pm 0.005 \arcdeg$ & $0.461 \pm 0.003 \arcdeg$ &
                $0.461 \pm 0.003 \arcdeg$ \nl
        & $y$ FWHM & $1.42 \pm 0.02 \arcdeg$ & $1.42 \pm 0.02 \arcdeg$ &
                $1.08\pm 0.02 \arcdeg$ & $0.513 \pm 0.004 \arcdeg$ &
                $0.513 \pm 0.004 \arcdeg$ \nl
Chopper & Pattern & Sine & Linear & Linear & Sine & Sine \nl
        & Amplitude on Sky & 4.90\arcdeg & 7.00\arcdeg & 7.35\arcdeg
                & 7.96\arcdeg & 3.36\arcdeg \nl
        & Rate & 3.906 Hz & 3.906 Hz & 3.906 Hz & 2.976 Hz & 2.976 Hz \nl
        & Samples/Sweep & 16 & 64 & 64 & 168 & 168 \nl
Pointing& Azimuth (W:E) & $-8.2\arcdeg$ : $7.8\arcdeg$ & $-7.15\arcdeg$ :
                $7.25\arcdeg$ & $-7.15\arcdeg$ : $7.25\arcdeg$ &
                $-7.392\arcdeg$ : $7.288\arcdeg$ & $0.14\arcdeg$ \nl
        & Jitter & \ldots & 0.02\arcdeg & 0.02\arcdeg & 0.03\arcdeg &
                0.03\arcdeg \nl
        & Elevation & 52.2\arcdeg & 52.2\arcdeg & 52.2\arcdeg &
                52.24\arcdeg & 60.2\arcdeg \nl
Timing  & Time Per Base Move & 16.4 s & 20 s & 20 s & 40.3 s & 2 weeks\nl
        & File Length & 17 min & 15 min & 15 min & 20 min & 20 min \nl

\enddata
 \end{deluxetable}

 The observing strategy is designed to offer a variety of internal
 systematic checks, minimize contamination by atmospheric temperature
 gradients, and to simultaneously probe a variety of angular scales.
 These goals are achieved by sweeping the beam in azimuth on the sky by
 many beam widths with a large chopping flat, and then synthesizing
 effective antenna patterns in software.  For simplicity in the
 discussion which follows, telescope and observing parameters for the
 1995 season are used.  Parameters for all years are listed in
 Table~\ref{table:obParms}.

 As the beam is swept on the sky, the radiometer is sampled 168 times
 per complete sweep.  In analysis the 168 samples from each sweep, $T_i$,
 are multiplied by a weighting vector, $w_i$, to give $\Delta T_{sweep} =
 \sum w_i T_i$.  By choice of weighting vector the relative
 weighting of each spatial point in the sweep can be set allowing the
 synthesis of arbitrary effective antenna sensitivity patterns,
 $H(x,y)$, given by
 \begin{equation} \label{eq:antenna}
   H({\bf x})=\left<\sum_{i} w_i G({\bf x}-
   {\bf X}_i)\right>_{bin}
 \end{equation}
 where ${\bf X}_i$ is the position on the sky of the center of the
 main beam corresponding to weighting vector element $w_i$ and
 \begin{equation}
   G(x,y) = \frac{1}{2\pi \sigma_x
     \sigma_y}\exp\left(-\frac{x^2}{2\sigma_x^2}
     -\frac{y^2}{2\sigma_y^2}\right)
 \end{equation}
 is the main beam pattern of the telescope pointed at $(0,0)$.  The
 azimuth dimension is denoted by $x$ and the elevation dimension by $y$.
 The beam width of the telescope is given by
 $\sigma=FWHM/\sqrt{8\ln(2)}$.  The effect of the rotation of the Earth
 during an integration is included in $H({\bf x})$, as
 denoted by $<\ldots>_{bin}$.

 For example, if the weighting vector $w_i$ is selected so that data
 which are acquired when the chopping plate is centered are given a positive
 weight, and data which are acquired when the chopping plate is offset to the
 left or right are given negative weight, then a three lobed synthesized
 beam on the sky is produced (hereafter referred to as the 3pt
 synthesized beam).  The weighting vectors are normalized so that
 $\int{|H({\bf x})|d{\bf x}}=2$ for a single sweep, not including
 the rotation of the Earth.  The mean of the weighting vector is zero in
 order to eliminate sensitivity to fluctuations in the mean temperature
 of the sky.  Since the weighting vectors for different effective
 antenna patterns are made to be nearly orthogonal all of the effective
 antenna patterns may be acquired from the same data.  Synthesized
 antenna patterns up to 19pt ($\approx 60$\,Hz) are produced.  The number
 of lobes is limited by the beam size to throw ratio, and not by the
 detectors.

 The synthesis of the 9pt beams is demonstrated in
 Figure~\ref{fig:synth}.  Figure~\ref{fig:antennas} displays some of the
 synthesized beams.  In total, including the three years, there are 37.
 The 3pt through 9pt antenna patterns are similar for \Ka94, Q94
 and Q95, allowing these data to be compared.  Since the sweep
 amplitude and pointing for SK93 was substantially different than for
 the other years, the \Ka93 3pt beam is also substantially different.
 Comparing SK93 with SK94 requires the use of a specially optimized
 weighting vector (\Ka93 overlap).

 Data are acquired in two modes: In the first mode (\CAP Data),
 which was used all three years, the chopping plate axis is vertical which
 places the beam at a constant elevation of $52.2\arcdeg$, the elevation
 of the North Celestial Pole (NCP) in Saskatoon.  The beam is swept
 sinusoidally $12.8\arcdeg$ in azimuth ($7.35\arcdeg$ peak to peak on
 the sky) at 2.976 Hz.  The center of the sweep is alternated every 40.3
 seconds between $7.32\arcdeg$ East of the NCP and $7.28\arcdeg$ West of
 the NCP.  As the earth rotates throughout the day, the celestial polar
 cap is covered, from the NCP to $82\arcdeg$ declination.  Data acquired
 in the East are repeated $\approx 12$ hours later in the West, giving a
 very powerful systematic test.  These data are multiplied by weighting
 vectors which produce a family of synthesized beams on the sky from 3pt
 to 19pt, and are then integrated into 24 (for 3pt to 5pt) or 48 (for 6pt
 to 19pt) bins in Right Ascension (RA).

 In the second mode (\RING Data), which was only used in 1995, the
 chopping plate is tilted back $4.0\arcdeg$ which raises the beam elevation
 $8.0\arcdeg$ to 60.2\arcdeg.  The telescope is pointed so that the
 center of the sweep is North.  The beam is swept sinusoidally with a
 peak to peak amplitude of 3.36\arcdeg on the sky.  In the small angle
 approximation, the beam is swept in RA at a constant declination.  As
 the Earth rotates an entire ring at $82\arcdeg$ declination is
 sampled.

 Synthesized beams are generated in the same way as with the \CAP
 data, except that a new weighting vector must be generated every
 sweep to keep the synthesized beam fixed in RA during its integration.
 The synthesized beams are smaller than the amplitude of the sweep which
 allows positions on the sky to be tracked in software.  Three point to
 6pt synthesized beams are produced, as well as the effective 3pt beam
 pattern of the \MS\ experiment.  The \RING data lacks the East to
 West comparison test of the \CAP data.  However, synthesis of the
 MSAM beam makes it possible to compare directly with a very different
 experiment.

\section{Data Selection and Reduction} \label{sec:data}

\begin{figure}[bp]
  \plotone{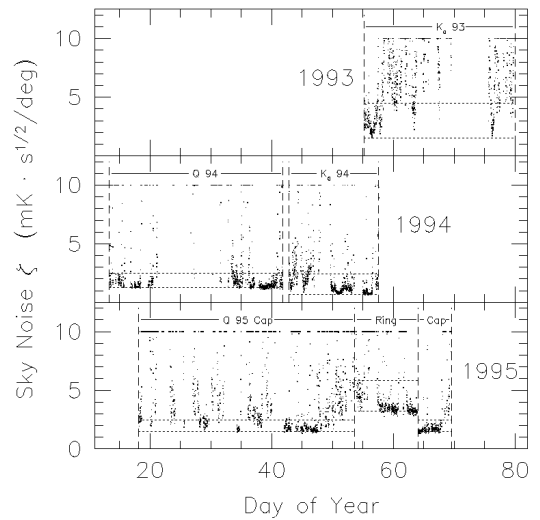}
  \caption{\small The sky noise in Saskatoon. The value $\zeta$ is a measure of the
    stability of the horizontal component of spatial temperature
    gradients (see text).  The upper dashed line in each plot denotes
    the nominal atmospheric cut.  The lower dashed line indicates the
    system noise limit, $\zeta_{n}=\kappa{\rm NET}/\theta_{eff}$.
    The noise floor for the 1995 \RING data is higher than for the
    \CAP data because $\theta_{eff}$ is smaller.  Data with
    $\zeta>10$ mK$ \cdot$ s$^{1/2}/\deg$ have been limited to $\zeta=10$
    mK$ \cdot$ s$^{1/2}/\deg$ in this plot.  The maximum value of
    $\zeta$ recorded was $1360$ mK$ \cdot$ s$^{1/2}/\deg$. }
  \label{fig:xi}
\end{figure}

\begin{table}[bp]
\begin{center}
\begin{tabular}{lccc}  
\hline\hline
        & $\kappa/\theta_{eff}$ & Cut Level $\zeta$       & hours \\
        & &(${\rm mK \cdot s^{1/2}}/\deg$) & \\
\hline
\Ka93   & 0.60  & 4.5  & 130    \\
        &       & 3.0  & 79     \\
\Ka94   & 0.44  & 2.5  & 140   \\
        &       & 1.7  & 107    \\
Q94     & 0.42  & 2.5  & 158    \\
        &       & 1.7  & 98     \\
Q95 \CAP & 0.36  & 2.5  & 174    \\
        &       & 3.5  & 242    \\
        &       & 1.9  & 114    \\
Q95 \RING& 0.86  & 6.8  & 121    \\
        &       & 9.7  & 132    \\
        &       & By Hand & 66  \\
\hline
\end{tabular}
\caption{Amount of data at different cut levels. The first cut
level for each experiment listed is the nominal cut used.  The others
are used for consistency checks.  $\kappa/\theta_{eff}$ is used to
convert $\zeta$ to receiver NET (see text). \label{table:xi}}
\end{center}
 \end{table}

 Because the SK telescope observes from the ground in Saskatoon, an
 important part of the analysis is the selection of data where
 contamination from atmospheric spatial temperature fluctuations do not
 dominate.  The first cut is made by evaluating, for each 20 minute
 file, the mean deviation of 8 second averages made with a beam
 synthesized using a 2pt weighting vector which is sensitive to the
 horizontal component of spatial temperature gradients.  This number is
 large when the spatial gradient is changing and small when it is
 stable and has been found to be a good indicator of atmospheric
 contamination.  The mean deviation is used because it is less sensitive
 than the standard deviation to spikes in the data, which are more
 likely to be due to birds or airplanes than to atmospheric noise.  The
 mean deviation is related to the standard deviation by
 $\Delta_x=\sqrt{\frac{2}{\pi}}\sigma_x$ for normal distributions.

 A more general expression for the cut levels is given by
 \begin{equation}
   \zeta=\sigma \sqrt{\tau}/\theta_{eff}
 \end{equation}
 where $\sigma$ is the standard deviation of the 2pt data,
 $\tau$ is the integration time, and
 \begin{equation}
  \theta_{eff}=\int{\theta_x H(\theta_x,\theta_y)d\theta_x d\theta_y}
 \end{equation}
 is the effective sweep amplitude in degrees.  The units of $\zeta$ are
 $mK\cdot s^{1/2}/\deg$.  Neglecting the radiometer contribution, this
 measurement of the stability of the atmosphere may be compared from
 experiment to experiment.  The effective sky noise, $\zeta_{n}$, due to
 the receiver NET is given by
 $\zeta_{n}\equiv\kappa{\rm NET}/\theta_{eff}$, where
 $\kappa=\sqrt{N\sum_i w_i^2}$, and $N$ is the number of samples per
 sweep.

 Figure~\ref{fig:xi} shows $\zeta$ as a function of time for each of the
 observing seasons.  Horizontal dashed lines are the nominal cut levels
 (the upper line) and the receiver noise level, $\zeta_{n}$ (the lower
 line), for each of the observing runs.  Each point represents the
 result for one file of data (see Table~\ref{table:obParms}).  Because
 the throw for the \RING data was significantly smaller than for
 the \CAP data, $\zeta_{n}$ is significantly higher.  In other
 words, the \RING observations are significantly less sensitive to
 the atmospheric gradients than the \CAP observations.

 It is furthermore required that for a file to be accepted, the files
 before and after must also pass the atmospheric cuts.  Thus, the sky
 must be stable for three files (45 minutes in 1994 and 60 minutes for
 1995) for any data to be accepted at all.  Of the data acquired, $27\%$
 passes the nominal cut.  Table~\ref{table:xi} summarizes the amount of
 data acquired in each observing mode at different cut levels.  The cut
 level listed first for each experiment is the one used for the
 definitive data set.  The lower cut data for the \RING is
 comprised of the two long receiver noise limited stretches around day
 59 and day 63 of 1995 rather than being based on specific sky noise
 level.  The analysis is performed for each of the listed cut levels
 (see Table~\ref{table:dTlG}).  Decreasing the cut level from the
 definitive values changes the spectrum by approximately $1\sigma$ for
 the SK95 data.  Because this lower cut level is so close to the receiver
 noise, good data are cut at random, reducing the length of contiguous
 stretches of data.  This increases sensitivity to long time scale
 noise and reduces the robustness of error bar calculation.  Increasing
 the cut level has an insignificant effect on the results.

 The data are blanked for base moves (e.g. 4 seconds every 40 seconds
 for SK95) and for spikes which may be due to birds, planes or
 occasional glitches in the data system.  An entire sweep of data is
 blanked if any of the samples in it (64 for SK94, and 168 for SK95)
 exceed the mean by more than some limit ($3.5\sigma$ for SK94 and
 $3.85\sigma$ for SK95).  The amount of data removed in this step is
 consistent with the data being normally distributed ($15\%$ for SK95).

 The data are then multiplied by the weighting vector and binned
 according to RA.  A weighted mean of the frequency and
 polarization channels is also made before application of the weighting
 vector for analyses where frequency and polarization information are not
 required.  For the \RING data, a new weighting vector is
 generated every sweep so that the antenna pattern in each bin is fixed
 in RA, while for the \CAP data, the weighting vectors are fixed.
 The \CAP data acquired in the East are binned separately from the data
 acquired in the West.  The bin numbering for both East and West data is
 based on the RA of the center of the sweep when the telescope is pointed in
 the East.

\begin{table}[tbp]
\begin{center}
\begin{tabular}{rcrr}  
\hline\hline
        & $\underline{\sigma_{15min}}$ & A Offset  & B Offset \\
        &       $\sigma_{18s}$          &  &  \\

\hline
\Ka93   3pt & 1.2    & $-60\pm 3\uK$    & $-360\pm 3\uK$ \\
\Ka94   3pt & 1.1    & $117\pm 6$    & $-550\pm 7$  \\
        4pt & 1.3    & $10\pm 5$     & $-50\pm 6$  \\
        5pt & 1.0    & $54\pm 6$     & $225\pm 6$  \\
        6pt & 1.0    & $1\pm 7$      & $5\pm 7$    \\
        7pt & 1.0    & $7\pm 9$      & $62\pm 9$   \\
        8pt & 1.0    & $0\pm 13$     & $34\pm 13$  \\
        9pt & 1.0    & $80\pm 19$    & $97\pm 18$   \\
Q94     3pt & 1.2    & $\ldots$      & $-695\pm 8$ \\
        4pt & 1.2    & $\ldots$      & $63\pm 7$ \\
        5pt & 1.0    & $\ldots$      & $281\pm 8$ \\
        6pt & 1.0    & $\ldots$      & $15\pm 8$ \\
        7pt & 1.1    & $\ldots$      & $79\pm 9$ \\
        8pt & 1.0    & $\ldots$      & $1\pm 11$ \\
        9pt & 1.0    & $\ldots$      & $51\pm 13$ \\
Q95     3pt & 1.5    & $-616\pm 13$   & $-2182\pm 9$ \\
        4pt & 1.4    & $4 \pm 11$    & $31\pm 7$ \\
        5pt & 1.2    & $89\pm 10$    & $436\pm 6$ \\
        6pt & 1.2    & $63\pm 9$     & $107\pm 6$ \\
        7pt & 1.1    & $21\pm 9$     & $145\pm 6$ \\
        8pt & 1.2    & $-56\pm 9$    & $-83\pm 6$ \\
        9pt & 1.1    & $21\pm 9$     & $53\pm 6$ \\
        10pt & 1.2   & $-101\pm 10$  & $-103\pm 6$ \\
        11pt & 1.1   & $25\pm 10$    & $35 \pm 6$ \\
        12pt & 1.1   & $70\pm 10$    & $75\pm 6$ \\
        13pt & 1.1   & $-26\pm 11$   & $7\pm 7$ \\
        14pt & 1.0   & $-34\pm 12$   & $-41\pm 7$ \\
        15pt & 1.1   & $-23\pm 12$   & $-31\pm 8$ \\
        16pt & 1.1   & $-12\pm 13$   & $-17\pm 8$ \\
        17pt & 1.0   & $15\pm 14$    & $12\pm 9$ \\
        18pt & 1.1   & $-41\pm 15$   & $-49\pm 9$ \\
        19pt & 1.0   & $-65\pm 17$   & $-31\pm 10$ \\

\hline
\end{tabular}
 \caption{Offsets and long time scale noise. 
   The ratio of error bars based on the
   distribution of 18s averages to error bars based on the distribution
   of 15 min averages is $\sigma_{15min}/\sigma_{18s}$.  The 15 minute
   averages are used.  {\it A Offset} is the offset in the vertically
   polarized channel.  {\it B Offset} is the offset in the horizontally
   polarized channel.  For Q94, the A receiver chain was damaged in
   shipping and was not used.
   \label{table:offsets}}
 \end{center}
 \end{table}

 For the \CAP data, the error bars are generated in two stages.
 First, after beam synthesis, the data are averaged into bins 15
 sidereal minutes wide.  Weights for each bin are calculated based on
 the distribution of 18 second averages.  If any 18 second average is
 more than $3.5\sigma$ away from the mean of the bin it is cut.  These
 15 min averages are then binned into 24 or 48 bins around the sky with
 weights propagated from the weights of the 15 minute averages.  These
 weights are used to give the relative weighting of each RA bin.  The
 overall weighting is generated from the distribution of the 15 minute
 bins after removal of corresponding mean signal from the 24 or 48 RA
 bins.  For data weighted to synthesize many lobed beams (6pt or more)
 the error bars based on 18 second averages agree to $10\%$ with those
 based on 15 minute averages, but for the 3pt or 4pt data (which is more
 sensitive to atmospheric noise) the distribution of 15 minute averages
 predicts error bars as much as $50\%$ larger than would be generated
 from the distribution of the 18 second averages (see
 Table~\ref{table:offsets}).  For the \RING data, with its
 substantially smaller sweep which yields much less sensitivity to the
 atmosphere, the error bars are generated directly from the distribution
 of the 18 second averages.

 \section{Offsets}

 The design of the SK telescope is intended to minimize the
 instrumental offset.  However, the emission from the large chopping
 plate is a function of the plate orientation, which theoretically
 produces a $\approx -400\uK$ offset for the horizontally polarized
 channel and a $\approx 200\uK$ offset in the vertically polarized
 channel in the 3pt data.  The measured 3pt offsets are approximately
 this magnitude for SK93 and SK94.  In SK95, however, an additional
 $\approx -1$\,mK offset was present in both polarizations.  The source
 of this additional offset is not known.  See \cite{Wol96} for
 possible explanations.

 The offsets are dealt with differently between the \CAP and
 \RING data analysis.  In the \CAP analysis, the weighting
 $w_i$ is constant for a given synthesized beam and chopping plate
 orientation.  For this reason the offsets are ignored until the data
 have been multiplied by the weighting vector and binned in RA.  At that
 point, the mean value of the 24 or 48 RA bins is removed.  The error
 bars are then multiplied by $\sqrt{24/23}$ or $\sqrt{48/47}$ to
 compensate for the removed degree of freedom.  The levels of the
 \CAP offsets for each synthesized beam are listed in
 Table~\ref{table:offsets}.

 For the \RING data, however, a new weighting vector must be
 calculated for each sweep, which causes the offset to change as the
 synthesized beam is tracked over the sky.  To deal with this, all of
 the accepted data are coadded synchronously with the sweep to form a
 168 sample mean sweep.  This mean sweep, which is essentially the
 instrumental offset as a function of chopping plate position, is
 subtracted from the data for each sweep prior to multiplication by the
 weighting vector.  At this point then, the offset for any synthesized
 \RING beam is zero.

 As evidenced by the larger in scatter of the 15 minute averages
 compared to that predicted by the distribution of the 18 second
 averages in the \CAP data (see Table~\ref{table:offsets}), there
 may be some drift to this offset on longer time scales.  Since at least
 some of offset is due to emission, the excess long time scale noise
 could be expected to correlate with the temperatures of the optical
 components.  The correlations are not significant ($\approx 1\sigma$)
 however, and regressing out the optics temperatures has no effect on
 the long time scale noise.  Since the gain of the amplifiers is a
 function of temperature the long time scale noise might be expected to
 vary with receiver temperature.  But, as with the optics temperatures,
 there is no significant correlation with receiver temperature, and
 regressing it out has no effect.  In the final analysis, nothing is
 regressed out.

 To further investigate the effect of the excess long time scale noise
 in the \CAP data, polynomials in time (linear to 9th degree) were fit
 out of the data.  The maximum slope from fitting a line in time to the
 data is 7\uK per day for the SK95 3pt data.  The typical slope is less
 than 1\uK per day and consistent with zero.  As with regressing out the
 auxiliary temperatures, removing lines or polynomials has no
 significant effect on the long time scale noise or on the final angular
 power spectrum.  In the final analysis, only the constant offset is
 removed.

 To test the possibility that the long time scale noise is due to
 instrumental drifts, data taken in the West can be subtracted from data
 taken in the East 40 seconds earlier.  Doing this decreases the
 differences between the short and long time-scale based error bars.
 This subtraction, however, has no significant effect on the spectrum,
 beyond increasing the error bars: $(E-W)/\sqrt{2}$ will have the same
 $rms$, as $E$ or $W$ by themselves, but now there is only one data set
 to do statistics on, rather than two.  It is also possible that the
 excess long time scale noise is due to atmospheric contamination: this
 was the interpretation taken by the OVRO Ring experiment (\cite{Rea89})
 for a similar effect.

\section{Internal Consistency Checks}

\begin{deluxetable}{rccccccccc}
\tablecaption{Internal Consistency Tests I: 3pt to 9pt \CAP \label{table:symmetries}}
\tablehead{
\colhead{} & \colhead{\Ka93 Overlap} & \colhead{3pt} & \colhead{4pt} &
\colhead{5pt} & \colhead{6pt} & \colhead{7pt} & \colhead{8pt} &
\colhead{9pt} }  
\startdata
\Ka93 Sum & 3.76 & \nodata & \nodata & \nodata & \nodata & \nodata & \nodata & \nodata \nl
        E$-$W & 1.17 & \nodata & \nodata & \nodata & \nodata & \nodata & \nodata & \nodata \nl
        A$-$B & 1.0  & \nodata & \nodata & \nodata & \nodata & \nodata & \nodata & \nodata \nl\nl

\Ka94 Sum & 5.20 & 4.59 & 2.16 & 2.92 & 1.47 & 1.40 & 1.13 & 1.03 \nl
        A$-$B & 0.79 & 0.55 & 0.67 & 0.70 & 0.85 & 0.94 & 0.88 & 1.18 \nl
        E$-$W & 0.76 & 0.88 & 0.83 & 0.70 & 1.17 & 1.06 & 1.39 & 0.59 \nl
 Quadrature & \nodata & 1.2  & 0.7  & 1.0  & 1.2  & 1.3  & \nodata & \nodata \nl
\nl 
   Q94 Sum  & 3.21 & 2.15 & 1.97 & 1.14 & 1.06 & 1.27 & 1.19 & 1.53 \nl
        E$-$W & 0.91 & 0.67 & 1.06 & 1.09 & 0.68 & 1.08 & 1.05 & 0.84 \nl
 Quadrature & \nodata & 0.7  & 1.1  & 1.0  & 1.1  & 1.4  & \nodata & \nodata
\nl \nl

   Q95 Sum  & \nodata & 2.64 & 1.98 & 3.27 & 2.49 & 2.49 & 2.69 & 2.37 \nl
        A$-$B & \nodata & 0.91 & 1.05  & 1.09  & 0.74 & 0.73 & 1.1  & 1.0 \nl
        E$-$W & \nodata & 1.13 & 0.99 & 1.22 & 1.12 & 1.60 & 1.09 & 1.16 \nl
      H1$-$H2 & \nodata & 2.34 & 1.60 & 1.05 & 1.13 & 1.09 & 1.06 & 0.88 \nl
 Quadrature & \nodata & 0.93 & 1.23 & 1.37 & 1.49 & 0.90 & 1.21 & 0.99 \nl
\nl 

\Ka93+\Ka94 & 6.12 & \nodata & \nodata & \nodata & \nodata & \nodata &
\nodata & \nodata \nl
\Ka93$-$\Ka94 & 1.18 & \nodata & \nodata & \nodata & \nodata & \nodata
& \nodata & \nodata \nl\nl
\Ka94+Q94 & 7.17 & 6.01 & 2.91 & 3.35 & 1.66 & 1.52 & 1.50 & 1.58 \nl 
\Ka94$-$Q94 & 1.25 & 0.73 & 1.22 & 0.71 & 0.87 & 1.15 & 0.82 & 1.03 \nl\nl
\Ka94+Q95 & \nodata & 6.01 & 2.94 & 5.21 & 2.53 & 2.50 & 2.66 & 2.56 \nl
\Ka94$-$Q95 & \nodata & 1.22 & 1.20 & 0.98 & 1.43 & 1.29 & 1.17 & 0.85 \nl\nl
Q94+Q95 & \nodata & 4.10 & 2.85 & 2.92 & 2.07 & 2.69 & 3.10 & 2.94 \nl
Q94$-$Q95 & \nodata & 0.69 & 1.09 & 1.49 & 1.48 & 1.06 & 0.78 &
1.02 \nl

\tablecomments{ Reduced $\chi^2$ for a variety of internal symmetry
  tests for 3pt to 9pt data.  The {\it Sum} entries indicate the
  reduced $\chi^2$ for the total signal on the sky (with A and B
  polarizations averaged and data acquired in the East and in the West
  averaged).  The \EW\ entries indicate the residual signal if data
  acquired in the West is subtracted from data acquired in the East 12
  hours earlier.  The \AB\ entries represent the residual if the two
  polarizations are differenced.  The {\it Quadrature} entries give the
  amplitude of the signal if the weighting vector is modified to have no
  sensitivity on the sky.  The {\it \Ka93 Overlap}, $3pt$, $4pt$ and
  $5pt$ entries have 23 degrees of freedom.  The rest have 47 degrees of
  freedom.  The {\it Sum} entries have consistently larger reduced
  $\chi^2$ entries than the the control entries, indicating the presence
  of a true sky signal.} \enddata
\end{deluxetable}

\begin{deluxetable}{rcccccccccc}
\tablecaption{Internal Consistency Tests II: 10pt to 19pt \CAP \label{table:symmetries2}}
\tablehead{
\colhead{} & \colhead{10pt} & \colhead{11pt} & \colhead{12pt} & \colhead{13pt} &
\colhead{14pt} & \colhead{15pt} & \colhead{16pt} & \colhead{17pt} & \colhead{18pt} &
\colhead{19pt} } 
\startdata
   Q95 Sum & 1.95 & 2.12 & 2.29 & 2.30 & 1.71 & 1.94 & 1.62 & 1.70 & 1.51 & 1.18 \nl
       A$-$B & 1.01 & 1.36 & 0.78 & 0.85 & 1.36 & 0.96 & 0.88 & 1.04 & 1.17 & 0.97 \nl
       E$-$W & 0.85 & 0.81 & 1.29 & 1.42 & 0.90 & 0.85 & 1.21 & 1.21 & 1.01 & 1.36 \nl
     H1$-$H2 & 0.91 & 1.01 & 1.29 & 0.91 & 1.53 & 1.62 & 0.73 & 1.29 & 1.19 & 1.57 \nl
Quadrature & 0.97 & 0.90 & 1.22 & 1.05 & 1.30 & 1.31 & 0.81 & 1.14 & 1.09 & 1.26 \nl
\tablecomments{
Reduced $\chi^2$ for a variety of internal symmetry tests for 10pt to 19pt
SK95 \CAP data.  All entries have 47 degrees of freedom.  
}
\enddata
\end{deluxetable}

\begin{deluxetable}{rccccc}
\tablecaption{Internal Consistency Tests III: \RING \label{table:symmetries3}}
\tablehead{
\colhead{} & \colhead{3pt} & \colhead{4pt} & \colhead{5pt} & \colhead{6pt} &
\colhead{MSAM Overlap}} 
\startdata
    Q95 Sum & 1.86 & 1.83 & 1.34 & 1.01 & 1.87 \nl
        A$-$B & 1.16 & 1.31 & 1.40 & 1.28 & 1.55 \nl
      H1$-$H2 & 1.19 & 1.07 & 1.24 & 1.2 & 1.05 \nl\nl
Q95+\MS & \nodata & \nodata & \nodata & \nodata & 3.43 \nl
Q95$-$\MS & \nodata & \nodata & \nodata & \nodata & 1.05 \nl
\tablecomments{
Reduced $\chi^2$ for a variety of internal symmetry tests for SK95
\RING data.  The $3pt$ to $6pt$ entries have 95 degrees of
freedom.  The {\it MSAM Overlap} entries have 80 degrees of freedom.  }
\enddata
 \end{deluxetable}

 The observing scheme used in the SK experiment has a number of internal
 consistency checks.  For the \CAP data, the rotation of the Earth
 causes the same sky as was observed with the telescope pointed to the
 East of the NCP to be observed $\approx 12$ hours later when the
 telescope is pointed to the West.  This test (\EW) gives
 assurance that the signal which has been observed is truly on the sky:
 contamination from the atmosphere or ground pickup could be expected to
 repeat on a 24 solar hour time scale, not a 24 sidereal hour time scale
 with the signal in the East lagged 12 sidereal hours from the signal in
 the West.

 The data set may also be divided into first half ($H1$) and second
 half ($H2$), and compared.  Since there is approximately one month
 between the centroids of the first and second half, the success of this
 test gives assurances that the observed signal repeats on sidereal
 time, and not solar time: if the signal repeated on solar time, the
 signal between halves would be lagged by 4 out of 48 bins.  This 
 places limits on contamination by the sun and any Radio Frequency
 Interference (RFI) that recurs daily.

 The data from the two polarization channels ($A$ is vertical and $B$ is
 horizontal) may also be compared.  This test is useful for detecting
 contaminants which effect the two receiver chains differently, such as
 data system pickup or polarized diffraction off of the ground screen.

 Since the same region of sky has been observed in 1994 as 1995, the 3pt
 to 9pt \CAP data can also be compared between years.  A
 comparison of the SK93 and SK94 data using a specially prepared
 weighting vector for SK94 is presented in Netterfield et al. (1995) and
 repeated here.  Since many elements of the telescope, radiometer, and
 ground shield were changed between years, this test can give assurances
 that the data have not been contaminated by side lobes, vibrations or
 atmospheric gradients.

 Finally, weighting vectors may be applied which have the same frequency
 as sky-sensitive weighting vectors, but with the phase adjusted
 to have no sensitivity to the sky (quadrature phase).  This test is
 sensitive to synchronous vibrational or electrical signals, which would
 not necessarily have the same phase as a true sky signal.

 The results of this set of internal consistency tests are given in
 Tables~\ref{table:symmetries}-\ref{table:symmetries3}.  The expected
 reduced $\chi^2$ for the difference tests should be distributed around
 1.0 with a standard deviation of $1/\sqrt{2\nu}$, where $\nu$ is the
 number of degrees of freedom, assuming that the sky signals are $100\%$
 correlated and the error bars are completely uncorrelated.  This is the
 case for the \HH\ tests, but not for the \EW\ tests, the \AB\ tests, or
 the year to year tests.  The small beam, large throw in azimuth (rather
 than declination) and slight asymmetry in the East to West pointing of
 the SK95 experiment means that the \EW\ lagged correlation is not
 $100\%$.  For the 12pt data, for instance, the theoretical \EW\ 
 correlation coefficient (see Section~\ref{sec:like}) is
 $\rho^{EW}_T\equiv C^{EW}_T/\sqrt{C^{EE}_T C^{WW}_T}= 0.75$.  For the
 3pt data, $\rho^{EW}_T=0.90$. In this case, where the sky signal is not
 $100\%$ correlated, we find\footnote{For two correlated zero mean normal
   deviates $\{{\bf x}\}$ and $\{{\bf y}\}$ with
   $\sigma^2_x=\sigma^2_y$, we write $y_i=a_i+\rho x_i$ where $\{{\bf
     a}\}$ is uncorrelated with $\{{\bf x}\}$ and $\rho$ is the
   correlation coefficient between $\{{\bf x}\}$ and $\{{\bf y}\}$.  One
   finds that
   $\sigma^2_{x+y}=(1+\rho)\sigma^2_x$ and
   $\sigma^2_{x-y}=(1-\rho)\sigma^2_x$.  After identifying $\{{\bf x}\}$
   with the East data and $\{{\bf y}\}$ with the West data, and noting
   that $\chi^2/\nu=1+\sigma^2_{x\pm y}/\sigma^2$ (where $\sigma$ is the
   instrument noise), we find equation (\ref{eq:EWcor}).}, 
 \begin{equation} \label{eq:EWcor}
 \chi^2_{E-W}/\nu=\frac{2\rho+(1-\rho)\chi^2_{E+W}/\nu}{1+\rho}
 \end{equation}
 For
 the 12pt data, this results in $\chi^2_{E-W}/\nu=1.19 \pm 0.1$ which is
 consistent with the value of 1.29 which was measured from the data.
 Similarly, atmospheric noise correlates the error bars from the $A$ and
 $B$ polarizations in the large angular scale data.  In this case we
 expect $\chi^2_{A-B}/\nu=(1-\rho^{AB}_D)$, where $\rho^{AB}_D\equiv
 C^{AB}_D/\sqrt{C^{AA}_D C^{BB}_D}$ is the noise correlation
 coefficient.

 While the majority of the entries in
 Tables~\ref{table:symmetries} to \ref{table:symmetries3} are consistent
 with the the signal being fixed on the sky, there are some entries
 which indicate trouble.  The worst entry is for Q95 3pt \HH.
 The value of the reduced $\chi^2$ is 2.34, which is inconsistent
 with the expected value of 1.0.  The \HH\ residual is dominated by
 long drifts ($\approx 12$\,hr), which is consistent with the earlier
 observation that the 3pt data are especially effected by excess long
 time scale noise.  However, the \EW\ tests, and more significantly,
 the year to year tests work well for this data set, indicating that if
 the entire season is used the effect of the drifts is averaged out.  A
 similar effect at a lower level is seen in the Q95 4pt data.  Without
 the ability to compare with the other seasons we would be much less
 sanguine regarding the quality of the Q95 3pt and 4pt data.

 Other notably inconsistent tests are the 14pt and 15pt \HH\
 tests.  For both of these the \EW\ tests are consistent
 with a true sky signal.  However, without a comparison between years,
 we have no explanation for this.  Despite these problems there is
 strong evidence that the measured signal is on the sky.

 \section{Likelihood Analysis} \label{sec:like}

 In the analysis of these data, two questions are asked: 1) is the
 spectral index of the fluctuations consistent with that of the CMB, and
 2) what is the angular power spectrum of the fluctuations.  Comparison
 of the data with theoretical predictions may be done by comparison with
 the angular power spectrum.  The determination of parameters (such as
 the amplitude of the fluctuations at various angular scales, and their
 spectral indices) are made using maximum likelihood tests.  

 The Likelihood (or the probability density of the data, assuming a
 theory) is defined as
 \begin{equation} \label{eq:likelihood}
   L=\frac{{exp}(-\frac{1}{2}{\bf t}^t{\bf M}^{-1}{\bf t})}
   {(2\pi)^{N/2}|{\bf M}|^{1/2}}
 \end{equation}
 where ${\bf t}$ is a data set and ${\bf M}= {\bf C}_D +
 {\bf C}_T$ is a corresponding covariance matrix.  ${\bf C}_D$ is
 the data covariance matrix, and ${\bf C}_T$ is the pixel to pixel
 theory covariance matrix which is a function of the model parameters.
 The model parameters are adjusted to maximize $L$.  For the Saskatoon
 analysis, the elements of the data vector $t_i$ correspond to data
 taken looking at a specified RA bin on the sky, with a specified
 synthesized antenna pattern, in a given frequency and polarization
 channel.

 Confidence intervals on a parameter $T$ are found from the distribution
 of $L(T)$.  The best value for the parameter is determined by finding
 $T_{max}$ where $L(T_{max})$ is maximized.  For positive definite
 quantities (such as the amplitude of the fluctuations) a non-zero
 detection is claimed when $L(0)/L(T_{max})>0.15$.  In this case the
 error bars are found by finding $T_-$ such that
 \begin{equation}
   \frac{\int_0^{T_-}{L(T)dT}}{\int_0^{\infty}{L(T)dT}} = 0.1587
 \end{equation}
 for the lower limit, and finding $T_+$ so
 \begin{equation}
   \frac{\int_0^{T_+}{L(T)dT}}{\int_0^{\infty}{L(T)dT}} = 0.8413
 \end{equation}
 for the upper limit.  When $L(0)/L(T_{max})<0.15$, the $95\%$ upper
 limit is found by finding $T'$ such that
 \begin{equation}
   \frac{\int_0^{T'}{L(T)dT}}{\int_0^{\infty}{L(T)dT}} =0.95.  
 \end{equation}
 For non positive definite quantities (such as spectral indices) the
 technique is similar, except that the lower limits of the above integrals
 are $-\infty$ rather than $0$.

\begin{deluxetable}{llc}
\tablecaption{The Correlations \label{table:correlations}}
\tablehead{
\colhead{Between} & \colhead{Source} & 
\colhead{Magnitude$\left(\frac{C_D^{ij}}{\sqrt{C_D^{ii}C_D^{jj}}}\right)$}} 
\startdata
Frequency Channels & HEMT correlations, Atmosphere & 0.2 - 0.7 \\
Polarization Channels & Atmosphere                 & 0.0 - 0.3 \\
East/West          & Atmosphere, Offset drift      & 0.0 - 0.2 \\
Q95 $n$pt and $(n+2)$pt & non-orthogonal weighting vectors & 0.0 - 0.2 \\
\tablecomments{
  A summary of the noise correlations included in the analysis and their
  sources.
}
 \enddata
 \end{deluxetable}

 The data covariance matrix, ${\bf C}_D$, describes the signal due to
 instrumental noise in the data.  The diagonal matrix elements are
 simply the variance of the data derived from the distribution of the 15
 minute averages as described in Section 4.  The off-diagonal elements,
 summarized in Table~\ref{table:correlations}, describe other
 instrumental correlations.  For instance, the noise between frequency
 channels in the same radiometer chain is correlated by gain
 fluctuations in the HEMT amplifiers (\cite{Jar96}).  Similarly,
 atmospheric noise introduces correlations between simultaneously
 acquired channels.  Since the weighting vectors for SK95 are not
 strictly orthogonal in time, there will be noise correlations between,
 for example, the 13pt data and the 15pt data.  The noise from data
 acquired in the East is also correlated with the noise from data
 acquired in the West 20 seconds earlier.  This is related to the excess
 long time-scale noise discussed previously, and may be due to the
 atmosphere, or drifting instrumental offsets.  The noise
 correlation coefficients are generated from the distribution of 18
 second synthesized beam averages for all simultaneously acquired data.
 All known correlations are included in the analysis.

\begin{figure}[tbp]
\plotone{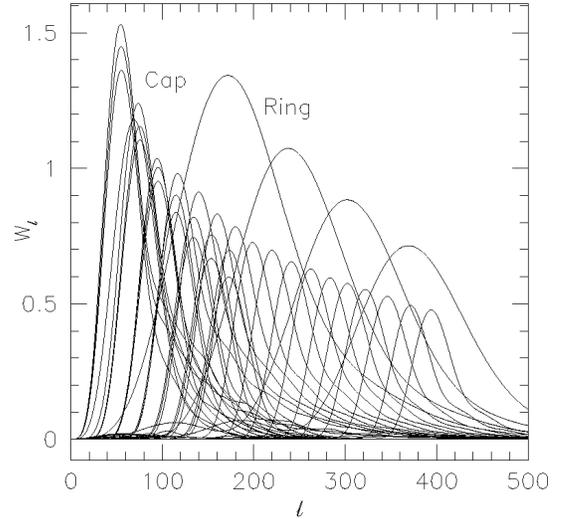}
\caption{The window functions for each synthesized beam for SK93-SK95.
  The many-lobed \CAP window functions are narrower than the
  \RING windows, providing finer angular resolution.}
\label{fig:windows}
 \end{figure}

 The theory covariance matrix, ${\bf C}_T$, describes the signal due
 to a hypothesized sky signal (\cite{Bon95} and references
 therein; \cite{Pee94}; \cite{Whi94}).  With the theoretical temperature
 fluctuations on the sky expressed in spherical harmonics as
 \begin{equation}
   T({\bf\hat{x}})=\sum_{\ell, m}a^m_\ell Y^m_\ell({\bf\hat{x}}),
 \end {equation}
 the theory covariance matrix can be expressed as
 \begin{equation} \label{eq:Ct}
   C_T^{ij}=\frac{1}{4\pi}\sum_{\ell}(2\ell+1)c_\ell W_\ell^{ij},
 \end{equation}
 where $c_\ell\equiv\left<|a_\ell^m|^2\right>$ and 
 \begin{equation} \label{eq:window}
   W_\ell^{ij}\equiv\int{d{\bf\hat{x}}_1}\int{d{\bf\hat{x}}_2
   H_i({\bf\hat{x}}_1)H_j({\bf\hat{x}}_2)
   P_\ell({\bf\hat{x}}_1\cdot{\bf\hat{x}}_2)}
 \end{equation}
 is the window function.  $H({\bf\hat{x}})$ is the effective antenna
 pattern from equation (\ref{eq:antenna}), and
 $P_\ell({\bf\hat{x}}_1\cdot{\bf\hat{x}}_2)$ are the Legendre polynomials.
 The window functions associated with the Saskatoon experiment
 are presented in Figure~\ref{fig:windows}.
 
 Rather than expressing the sky fluctuations in terms of $c_\ell$, we
 use
 \begin{equation} \label{eq:dTldef}
   \delta T_\ell \equiv \sqrt{\frac{\ell(2\ell+1)}{4\pi}c_\ell}.
 \end {equation}
 In terms of $\delta T_\ell$ the theory covariance matrix becomes
 \begin{equation} \label{eq:CtdTl}
 C_T^{ij}=\sum_{\ell} \delta T_{\ell}^2\frac{W_\ell^{ij}}{\ell}.
 \end{equation}
 So $\delta T_\ell^2$ is the variance per logarithmic interval of the angular
 spectrum of the CMB.  
 \section{The Window Function} 

 The angular scale to which a data set is sensitive ($\ell_e$) is
 determined from the diagonal components of the window function,
 $W_\ell^{ii}$ as follows. The expected {\it rms} amplitude of the
 fluctuations, $\Delta^i$, predicted by a theory for data acquired with a
 specified beam is given by the theoretical correlation matrix
 (eq. [\ref{eq:CtdTl}]) as
 \begin{equation}
 \Delta^i=\sqrt{C_T^{ii}}=\sqrt{\sum_{\ell} \delta T_\ell^2
 \frac{W_\ell^{ii}}{\ell}},
 \end{equation}
 or the sum of the theoretical angular spectrum, weighted by
 $W_\ell^{ii}/\ell$. Under the assumption of a flat spectrum in $\delta
 T_\ell$, this may be simplified to
 \begin{equation} \label{eq:rms}
 \Delta^i= \delta T_{\ell_e}\sqrt{I(W_\ell^{ii})}
 \end{equation}
 where
 \begin{equation}
 I(W_\ell) \equiv \sum_{\ell}\frac{W_\ell}{\ell}. 
 \end{equation}
 This provides a convenient conversion between the $rms$ amplitude of the
 fluctuations, $\Delta$, and the flat band power estimation, $\delta
 T_{\ell_e}$.  The effective $\ell$ for a single window function is the
 centroid of $W_\ell^{ii}/\ell$, or
 \begin{equation} \label{eq:le}
 \ell_e^{i}=\frac{I(\ell W_\ell^{ii})}{I(W_\ell^{ii})}.
 \end{equation}

 The band power estimate, $\delta T^i_{\ell_e}$, over a window function,
 $W_\ell^{ii}$, can be generated directly from the angular spectrum,
 $\delta T_\ell$, by
 \begin{equation} \label{eq:dTle}
   \delta T^2_{\ell_e^{i}}=\frac{I(\delta T_\ell^2
       W_\ell^{ii})} {I(W_\ell^{ii})}.
 \end{equation}
 When combining the results of several measurements with differing
 window functions it is appropriate to find the weighted mean of $\delta
 T^2_{\ell_e^{i}}$, since these are proportional to the variances, which
 are normally distributed far large degrees of freedom.  Thus,
 \begin{equation} \label{eq:dTle_}
   \overline{\delta T}_{\ell_e}^2 = \frac{\sum_i{\delta
       T_{\ell_e^{i}}^2/\sigma^2_{\delta T_{\ell_e^{i}}^2}}}
   {\sum_i{1/\sigma^2_{\delta T_{\ell_e^{i}}^2}}} 
 \end{equation}
 Inserting equation (\ref{eq:dTle}) into equation (\ref{eq:dTle_}) we find
 \begin{equation}
   \overline{\delta T}_{\ell_e}^2 = \sum_\ell{\delta T^2_\ell
     \frac{\overline{W}_\ell}{\ell}}
 \end{equation}
 where we have introduced the definition
 \begin{equation} \label{eq:sumWin}
   \overline{W}_\ell \equiv
   \frac{\sum_i{W_\ell^{ii}/\left(I(W_\ell^{ii})\sigma^2_{\delta
           T_{\ell_e^{i}}^2}\right)}} {\sum_i{1/\sigma^2_{\delta T_{\ell_e^{i}}^2}}}.
 \end{equation}
 $\overline{W}_\ell$ is the effective window function which results from
 combining band power measurements with different window functions.  The
 quantity $\sigma_{\delta T_{\ell_e^{i}}^2}$ is the measurement
 uncertainty on $\delta T_{\ell_e^{i}}^2$.  Note that
 $I\left(\overline{W}_\ell\right)\equiv 1$ with this definition.

 \section{The Theoretical Power Spectrum} \label{sec:dTl}

 If the angular spectrum varies slowly over the angular scales to
 which the experiment is sensitive it is reasonable to expand $\delta
 T_\ell$ as a power law in both angular scale, $\ell$, and observing
 frequency, $\nu$, as
 \begin {equation} \label{eq:dTlParam}
 \delta T_\ell = \delta T_{\ell_e} \left(\frac{\ell}{\ell_e}\right)^m
 \left(\frac{\nu}{\nu_o}\right)^\beta,
 \end {equation}
 where $\delta T_{\ell_e}$ is the amplitude of the fluctuations at
 angular scale $\ell_e$ and frequency $\nu_o$.  The angular spectral
 index is $m$ and the frequency spectral index is $\beta$.  With
 $\ell_e$ chosen using equation (\ref{eq:le}) and $\beta$ chosen as the center
 frequency of the observations (i.e.,36\,GHz) the three parameters are
 roughly orthogonal for $m$ and $\beta$ near $0$.  Therefore, neither
 the amplitude of the fluctuations, $\delta T_\ell$, nor the frequency
 spectral index, $\beta$ will be a function of choice of angular
 spectral index, $m$.

 Determination of the frequency spectral index, $\beta$ is useful for
 discriminating the source of the fluctuations.  For the CMB $\beta=0$,
 while in the Rayleigh-Jeans region for Galactic free-free emission
 $\beta=-2.1$, and for Galactic dust $\beta=1.7$.

 By combining data from multiple window functions the angular spectral
 index, $m$, may be determined.  Since the limits on this
 parameter are often uninterestingly large, it is often simply set to
 some fixed value.  Choosing $m=0$ assumes constant $\delta T_\ell$ over
 the region where the data are sensitive.  The determination of $\delta
 T_{\ell_e}$ for $m=0$ is known as a {\it flat band power} estimate of
 the angular spectrum.  The choice of $m=1$ reproduces the ``delta
 function'' correlation function used in \cite{Wol93} and \cite{Net95}.

 The procedure for finding the angular spectrum, $\delta T_\ell$, is to
 group data sensitive to a particular range of angular scales (e.g., the
 5pt data are sensitive to angular scales around $\ell=100$) and then
 use the maximum likelihood test to find the best amplitude, $\delta
 T_{\ell_e}$, and where possible the spectral indexes, $\beta$ and $m$,
 at each angular scale.  Where there is insufficient leverage to
 determine $\beta$ and $m$, they are fixed at 0.

 Of mainly historical interest is the Gaussian autocorrelation function,
 \begin{equation} \label{eq:gacf}
 \footnotesize
 C_T^{ij}(\theta_{12})=C(0)\int{d{\bf\hat{x}}_1}\int{d{\bf\hat{x}}_2
 H_i({\bf\hat{x}}_1)H_j({\bf\hat{x}}_2)
 \exp\left(\frac{-\theta_{12}^2}{2\theta_c^2}\right)},
 \end{equation}
 where
 $\theta_{12}={\rm acos}({\bf\hat{x}}_1\cdot{\bf\hat{x}}_2)$.
 The angular spectrum associated with this is
 \begin{equation}
 \delta T_\ell = \sqrt{C(0)\frac{\ell(2\ell+1)}{2\ell_o^2}
         \exp\left(-\frac{\ell^2}{2\ell_o^2}\right)}
 \end{equation}
 with $\ell_o=1/\theta_c$.  The coherence angle of the theory is
 $\theta_c$, and is often varied to maximize overlap with the
 experiment.  The free parameter which is varied in a maximum
 likelihood test is $C(0)$.  This spectrum has no particular theoretical
 motivation.  However, since it changes only slowly over the window
 function of most experiments, $C(0)$ can be converted to a band power,
 $\delta T_{\ell_e}$ by (\cite{Bon94})
 \begin{equation} \label{eq:C0_dTl}
 \delta T_{\ell_e} = \sqrt{C(0)\frac{I[u^2\exp(-u^2/2)W_\ell]}{I(W_\ell)}}
 \end{equation}
 where
 \begin{equation}
 u=\frac{\ell + 1/2}{\ell_o + 1/2}.
 \end{equation}

 \section{The {\it rms} Amplitude Analysis} \label{sec:rms}

 It is sometimes convenient to characterize the fluctuations by their
 {\it rms} amplitude, $\Delta_{sky}$, given by
 $\Delta^2_{sky}=\Delta^2_{total}-\Delta^2_{inst}$.  The raw {\it rms}
 of the data including instrument noise is
 $\Delta_{total}=\sqrt{\sum{\Delta T_i^2/N}}$, and the estimated
 contribution due to instrument noise is $\Delta_{inst}$.
 
 The instrument noise, $\Delta_{inst} \pm \sigma_{\Delta_{inst}}$, is
 generated from the data covariance matrix, $C_D$; 4096 examples of data
 sets described by $C_D$ are generated by
 \begin{equation}
   {\bf t} = C_D^{1/2}{\bf N}
 \end{equation}
 where ${\bf N}$ is a vector of the same dimension as ${\bf t}$
 comprised of samples drawn from a unit normal distribution.  The
 quantities $\Delta_{inst}$ and $\sigma_{\Delta_{inst}}$ are found from
 the distribution of the variances of the fake data sets.  The
 uncertainty in $\Delta_{sky}$ due to measurement noise is given by
 \begin{equation}
   \sigma_{\Delta_{sky}}^{inst}=
   \frac{\Delta_{inst}}{\Delta_{sky}}\sigma_{\Delta_{inst}}.
 \end{equation}
 
 This determination of $\Delta_{sky} \pm \sigma_{\Delta_{sky}}^{inst}$
 gives the amplitude (and uncertainty) of the fluctuations only over the
 region of sky which was measured, and is therefore only an estimate of
 the amplitude of the overall distribution from which this sample was
 drawn.  This is appropriate for comparing different measurements of the
 same sky but does not include sample variance.

 The uncertainty in $\Delta_{sky}$ due to sample variance
 ($\sigma^{SV}_{\Delta_{sky}}$) is generated from a theory covariance
 matrix, $C_T$, in the same way as $\sigma_{\Delta_{inst}}$ was generated
 from $C_D$; 4096 fake data sets described by $C_T$, are generated by
 \begin{equation}
   {\bf t} = C_T^{1/2}{\bf N}.
 \end{equation}
 The uncertainty $\sigma_{\Delta_{sky}}^{SV}$ is found from the
 distribution of the variances of the fake data sets.  This estimate of
 the uncertainty due to sample variance is larger than what is found by
 a Likelihood analysis; the latter includes the effect of spatial phase
 information inherent in the model which the former does not
 (\cite{Kno95}).  The uncertainties $\sigma^{SV}_{\Delta_{sky}}$ and
 $\sigma^{inst}_{\Delta_{sky}}$ are added in quadrature to estimate the
 total uncertainty in $\Delta_{sky}$.  The {\it rms} amplitude,
 $\Delta_{sky}$, and its uncertainties may be converted to $\delta
 T_\ell$ using equation (\ref{eq:rms}).

 \section{Determination of \\ 
   The Frequency Spectral Index $\beta$} \label{sec:beta}
 
 The first issue addressed in the analysis is whether the signal is due
 to the CMB or is due to some other foreground contaminant.  As noted,
 the spectral index of most foregrounds is expected to deviate from that
 of the CMB, $\beta=0$.  Since data are acquired in 6 frequency bins
 between 26\,GHz and 46\,GHz, $\beta$ for the fluctuations may be
 evaluated using the likelihood analysis described.  Since the data
 acquired with the \Ka-band radiometer only has adequate sensitivity
 for the 3pt, 4pt, and 5pt synthesis, this may only be done
 internally for angular scales below $\ell\approx100$.  The frequency
 span given by the Q-band data alone is inadequate to determine $\beta$.

 To evaluate $\beta$ the data from \Ka93 3pt and SK94 \& SK95
 3pt to 5pt are used.  The covariance matrix, {\bf M} in
 equation (\ref{eq:likelihood}) has dimension\footnote{ \Ka93 has 3 frequency
   channels, 21 RA bins, and both East and West data (126 rows), \Ka94,
   Q94 and Q95 each have 3 frequency channels, 24 RA bins, and East and
   West data for 3pt, 4pt and 5pt data (1296 rows).  } $1422 \times
 1422$.  For the theoretical spectrum (eq. [\ref{eq:dTlParam}]) is
 used with $m=0$.  The likelihood analysis yields $\delta T_{\ell_e} =
 47^{+7}_{-5}\uK$ and $\beta = 0.2^{+0.3}_{-0.3}$ for $\ell_e = 73$.
 Repeating the analysis with $m=-1$ yields $\beta = 0.2^{+0.3}_{-0.3}$,
 and $m=+1$ yields $\beta = 0.1^{+0.3}_{-0.3}$.  Thus the limits on
 $\beta$ are independent of choice of angular spectral index, $m$.  The
 channel to channel calibration uncertainty is $2\%$ to $5\%$.  This
 causes an additional uncertainty in $\beta$ of $\pm0.1$, which is to be
 added in quadrature to the quoted errors.
 
 The spectral index, $\beta$, of the fluctuations is consistent with
 that of the CMB ($\beta=0$), and inconsistent with known potential
 foreground contaminants.  In particular, contamination by Galactic
 synchrotron ($\beta=-2.8$), Galactic free-free emission ($\beta=-2.0$)
 and Galactic dust ($\beta=1.7$) in this data set are ruled out at large
 angular scales.  Foregrounds are discussed in detail in
 Section~\ref{sec:for}.

\section{Comparison With \MS}

Perhaps the most powerful systematic check available with the Saskatoon
experiment is the comparison with the \MS\ experiment.  \MS\ is a
balloon born experiment which observed the sky with a 3pt beam at a
declination of $82\arcdeg$ in $\approx 81$ fields between 14.4\,hr and
20.4\,hr RA (\cite{che94}).  While the Saskatoon experiment observes the
sky between 26\,GHz and 46\,GHz, the \MS\ experiment has observing bands
sensitive to the CMB at 180\,GHz and 240\,GHz, providing very wide
frequency coverage between the two experiments.  The comparison of the
two experiments of very different nature places stringent limits on
systematic errors or foreground contamination.

\begin{figure}[tbp]
\plotone{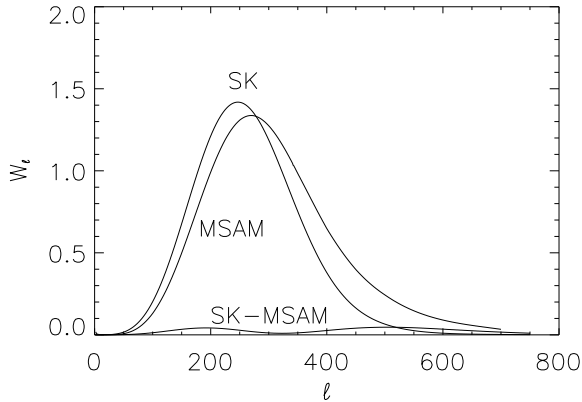}
\caption{The \MS\ and \MSOL\ window functions.  The window
functions are slightly different due to unequal beam size.  The window
function for the difference is also presented.} \label{fig:MSAMwin}
\end{figure}

 As mentioned in Section~\ref{sec:beams}, data from 1995 which was
 acquired in the \RING mode (at a declination of $82.05\arcdeg$)
 was multiplied by a weighting vector which closely synthesized the
 MSAM 3pt beam.  To generate this weighting vector, the elements $w_i$
 are adjusted to minimize the variance between the synthesized SK and
 MSAM beam patterns, assuming that the declination of the two
 experiments is equal.  In reality, the mean pointing of the two
 experiments differs by $0.14\arcdeg$, and the \MS\ experiment did
 not track in fixed elevation.  This discrepancy is ignored in the in
 the analysis of $\Delta$, but included in the Likelihood analysis.
 Once a SK weighting vector has been produced which maximizes spatial
 overlap, it is scaled so that $\sqrt{I(W)}=1.15$, the value for the
 MSAM window function.  To check the level of overlap between the the
 two experiments, the window function of the difference of the
 synthesized beams is formed.  For the difference beam
 $\sqrt{I(W)}=0.25$.  Given this overlap, it is expected that over
 $90\%$ of the sky signal will be in common between the two experiments.
 The window functions for the MSAM overlap are given in
 Figure~\ref{fig:MSAMwin}.

\begin{figure}[tbp]
  \plotone{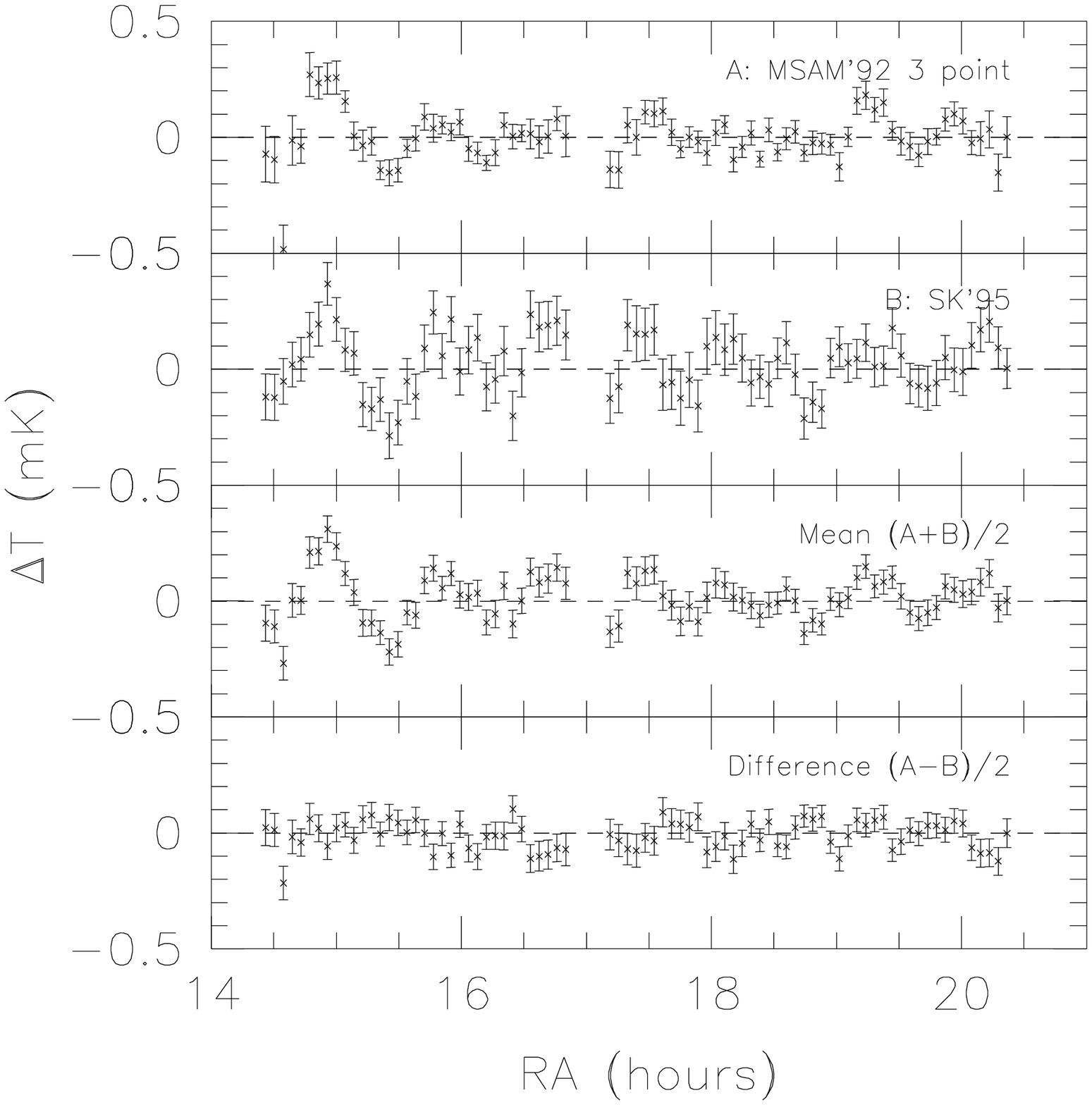}
  \caption{\MS\ and \MSOL\ compared.  The first panel
    gives the \MS\ 3pt data.  The second panel gives \MSOL\ 
    data for the same region.  The third shows the mean and the
    fourth the difference.  The data are in 85 equally spaced RA bins
    (only 81 have data), and are given in thermodynamic temperature
    units.  The anomalous third bin in the MSAM data set is dropped
    in the analysis of the comparison.  The $0.5\arcdeg$ FWHM beam spans
    0.24\,hr of RA.}
  \label{fig:MSAM}
\end{figure}

 Given the high degree of overlap expected, a naive test is adequate to
 verify that the two experiments see the same signal.
 Figure~\ref{fig:MSAM} shows the two data sets, their mean and
 difference.  The reduced $\chi^2$ for the sum is 3.43, while the
 reduced $\chi^2$ for the difference is 1.05, neglecting the anomalous
 third sample at $14.58$\,hr in the \MS\ data set.  It is clear that
 both experiments see the same signal on the sky.

 To find the consistency of the amplitudes measured by the two
 experiments, an analysis of $\Delta$ as described in
 Section~\ref{sec:rms} is performed.  The data covariance matrix for the
 \MS\ data ($C_{D_{MS}}$) is taken to be diagonal, neglecting
 correlations which were introduced by the removal of offset drifts
 in the data reduction.  For the SK data, $C_{D_{SK}}$ is made from the
 distribution of the 18 second averages, and includes all known
 correlations.  For analysis of the combined \MS/SK95 data, the
 weighted mean is used, and the off diagonal elements on the covariance
 matrix are generated using
 \begin{equation} 
   C_D^{ij}=C_{D_{SK}}^{ij} \frac{w_{SK}^{i}w_{MS}^{j}}
   {(w_{SK}^{j}+w_{MS}^{i})(w_{SK}^{i}+w_{MS}^{j})}
 \end{equation}
 where $w_{SK}^{i}$ and $w_{MS}^{j}$ are the statistical weights of the
 SK and \MS\ data respectively.  

 The results of this analysis are presented in Table~\ref{table:MSAM}.
 The errors on $\Delta$ include only $\sigma_{\Delta_{sky}}^{inst}$ and
 do not include sample variance.  This is appropriate for comparing the
 consistency of two measurements of the same sky.  The amplitudes are
 consistent, with SK favoring a higher value than MSAM at the $1\sigma$
 level. 

\begin{deluxetable}{rllll}
  \tablecaption{\MS\ and SK compared \label{table:MSAM}}
  \tablehead{
    \colhead{} & \colhead{\MS} & \colhead{SK} & \colhead{Mean} &
    \colhead{Difference}} 
  \startdata
  $\chi^2/\nu$ & 2.66 & 1.82 & 3.43 & 1.05 \nl
  $\Delta_{tot}$  & 88.9\uK & 104.4\uK & 76.6\uK & 52.2\uK \nl
  $\Delta_{inst} \pm \sigma_{\Delta_{inst}}$ & $70.5\pm5.3 \uK$ 
                              & $79.5\pm7.9 \uK$
                              & $49.3\pm3.9\uK$ & $53.2\pm4.3\uK$  \nl
  $\Delta_{sky} \pm \sigma_{\Delta_{inst}}^{inst}$ & $53.5\pm7.0\uK$
                              & $65.9\pm10.2\uK$ & 58.6$\pm3.3\uK$ 
                              & $\ldots$ \nl
  $\delta T_\ell(\Delta_{sky})$ & 46.5\uK & 57.3\uK & 51\uK & $\ldots$ \nl
  $\delta T_\ell$(Full Likelihood) & $52^{+11}_{-6}\uK$ 
                              & $61^{+19}_{-11}\uK$
                              & $50.9^{+10}_{-6} \uK$ & $\ldots$ \nl
 \tablecomments{A comparison of the \MS\ and \MSOL\ data.  The
  reduced $\chi^2$ entries are for the 80 bins in Figure~\ref{fig:MSAM},
  neglecting the anomalous third sample from \MS.  The uncertainties
  on $\Delta_{sky}$ do not include sample variance.  The uncertainties
  on $\delta T_\ell$ do.  $\delta T_\ell(\Delta_{sky})$ is $\delta
  T_\ell$ inferred from $\Delta_{sky}$ using equation (\ref{eq:rms}).
  Calibration uncertainties are not included.}

 \enddata
 \end{deluxetable}

 This analysis has not included calibration uncertainty in the two
 instruments.  For Saskatoon it is 14\%, and for \MS\ it is
 10\%.  Adding these in quadrature yields a relative calibration
 uncertainty of 17\%.  A least-squares fit to find the best relative
 calibration between the two data sets finds $N_{SK}/N_{MSAM} = 0.82 \pm
 0.16$.  The relative calibration based on the data favors a smaller
 normalization for SK or a larger normalization for MSAM, but is still
 consistent with the absolute calibrations used.

 The flat band power is found by a full likelihood analysis using the
 angular spectrum in equation (\ref{eq:dTlParam}).  These results
 are presented in Table~\ref{table:MSAM}.  The results for the two
 experiments are fully consistent.  Note that calibration uncertainties
 have not been included in quoted errors.

 The original \MS\ analysis is done in terms of the GACF, equation
 (\ref{eq:gacf}).  This analysis is repeated here.  For \MS, we find
 $\sqrt{C(0)}=70^{+15}_{-9}\uK$, using $\theta_c=0.92$, which is
 consistent with the analysis quoted in Cheng et al. (1994).
 (1994)\footnote{Cheng et al. (1994) find $50\uK<\sqrt{C(0)}<90\uK$ for
   the $5\%$ to $95\%$ confidence interval.  Reanalyzing the \MS\ data, we
   find $55\uK<\sqrt{C(0)}<99\uK$.} For SK we find
 $\sqrt{C(0)}=81^{+26}_{-15}\uK$.  Converted to band power estimates
 using equation (\ref{eq:C0_dTl}) we find 53\uK and 61\uK for \MS\ and SK
 respectively.

 A determination of $\beta$ of the fluctuations yields
 $\beta=-0.1\pm0.2$.  The combination of the SK95 and \MS\ 
 experiments has essentially measured the frequency spectrum of the
 anisotropy of the fluctuations from 36\,GHz to 240\,GHz and have found
 them to have the frequency spectrum of the CMB.  Over this large
 frequency range no other known source has this spectrum.  Free-free
 emission has $\beta=-1.45$ (as compared to -2.1 in the Rayleigh-Jeans
 region) and interstellar dust has $\beta=2.25$ (1.7 in the
 Rayleigh-Jeans region).  The spectral index of these fluctuations is
 highly inconsistent with these foreground contaminants.

 The differences between the two experiments render it very unlikely
 that the observed signal is the result of mutually common
 contamination.  MSAM is a balloon-borne experiment which observed the
 sky for several hours, while the Saskatoon experiment is ground based
 and observed the sky over many days.  MSAM is a bolometer based
 experiment which observes the sky at 180\,GHz and 240\,GHz while SK95
 is based on 40\,GHz HEMT amplifiers.  MSAM utilizes multi-mode optics
 while those of the Saskatoon telescope are single mode.  MSAM
 calibrates on Jupiter while SK calibrates on Cas-A.  The consistency of
 the measurements between MSAM and SK indicates that both experiments
 have measured the anisotropy of the CMB.

\section{Determination of \\ The Angular Power Spectrum}

\begin{figure}[tbp]
\plotone{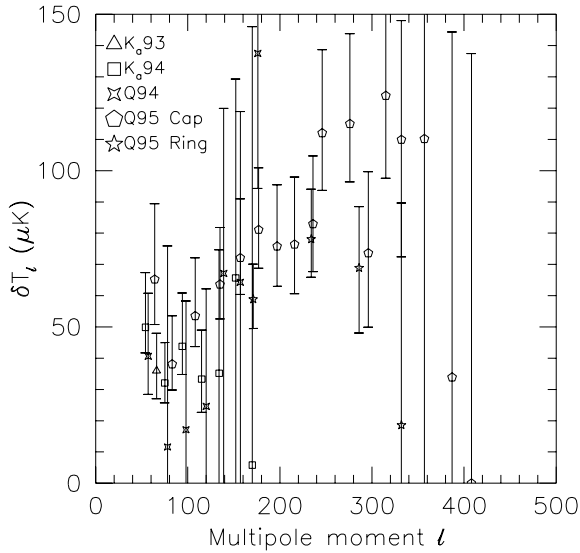}
\caption{The angular spectrum in 37 bins.  The results for each
  synthesized beam for SK93-SK95 is presented separately.  These results
  are given in Table~\ref{table:dTlsep}. The error bars do not include
the $\pm 14\%$ $(1\sigma)$ overall calibration
uncertainty.} \label{fig:dTlsep}
\end{figure}

\begin{deluxetable}{rcccccccccl}
\tablecaption{The angular spectrum in 37 bins
\label{table:dTlsep}}
\newlength{\pbwidth}
\tablehead{
\colhead{} & \colhead{Bins} & \colhead{$\Delta_{total}$} & 
\colhead{$\Delta_{inst}$} & \colhead{$\Delta_{sky}$} &
\colhead{$\sqrt{I(W)}$} & 
\colhead{$\delta T_\ell(\Delta_{sky}$)} &
\colhead{$\sigma^{inst}_{\delta T_\ell}$} & 
\colhead{$\sigma^{SV}_{\delta T_\ell}$} &
\colhead{$\ell_e$} &
\colhead{$\delta T_\ell$}
}
\startdata
\Ka93 3pt & 24 & 49\uK & 37\uK & 32\uK & 1.04 & 31\uK & 5\uK & 11\uK &
        $66^{+28}_{-19}$ & $36^{+12}_{-9}\uK$ \nl
\Ka94 3pt & 24 & 65\PuK & 44\PuK & 49\PuK & 1.16 & 42\PuK & 4\PuK &
        12\PuK & 
        $56^{+21}_{-18}$ & $50^{+18}_{-8}\PuK$ \nl
      4pt & 24 & 48\PuK & 42\PuK & 24\PuK & 0.88 & 27\PuK & 9\PuK &     
        11\PuK &
        $76^{+21}_{-19}$ & $32^{+13}_{-6}\PuK$ \nl 
      5pt & 24 & 42\PuK & 33\PuK & 26\PuK & 0.73 & 36\PuK & 10\PuK &
        10\PuK &
        $96^{+21}_{-19}$ & $44^{+17}_{-9}\PuK$ \nl
      6pt & 48 & 65\PuK & 56\PuK & 33\PuK & 0.62 & 54\PuK & 12\PuK & 
        9\PuK & 
        $115^{+21}_{-19}$& $33^{+16}_{-11}\PuK$ \nl
      7pt & 48 & 81\PuK & 75\PuK & 30\PuK & 0.54 & 56\PuK & 27\PuK & 
        13\PuK &
        $134^{+22}_{-19}$& $35, \ < 75$ \nl
      8pt & 48 & 111\PuK & 101\PuK & 47\PuK & 0.48 & 47\PuK & 36\PuK &
        12\PuK &   
        $152^{+23}_{-18}$ & $66, \ <129$ \nl
      9pt & 48 & 127\PuK & 146\PuK &  0\PuK & 0.43 &  0\PuK & \ldots\PuK &
        12\PuK  &   
        $170^{+24}_{-17}$ & $6, \ <146$ \nl
Q94  3pt & 24 & 86\PuK  & 73\PuK  & 45\PuK & 1.22 & 47\PuK & 10\PuK &
        11\PuK &
        $59^{+18}_{-21}$ & $41^{+20}_{-12}\PuK$ \nl   
     4pt & 24 & 83\PuK  & 65\PuK  & 53\PuK & 0.93 & 57\PuK & 9\PuK  &
        10\PuK &
        $81^{+16}_{-24}$ &   $12, \ <76$ \nl   
     5pt & 24 & 59\PuK  & 55\PuK  & 22\PuK & 0.77 & 28\PuK & 19\PuK &
        10\PuK &
        $100^{+17}_{-23}$ &   $17, \ <58$ \nl   
     6pt & 48 & 75\PuK  & 82\PuK  & 0\PuK  & 0.67 & 0\PuK  & \ldots\PuK &
        9\PuK &
         $120^{+17}_{-24}$ &   $25, \ <62$ \nl   
     7pt & 48 & 109\PuK & 99\PuK  & 45\PuK & 0.59 & 77\PuK & 27\PuK &
        12\PuK &
        $139^{+18}_{-24}$ &   $67, \ <120$ \nl   
     8pt & 48 & 112\PuK & 107\PuK & 32\PuK & 0.53 & 61\PuK & 50\PuK &
        11\PuK &
        $157^{+19}_{-23}$ &   $64, \ <119$ \nl   
     9pt & 48 & 136\PuK & 129\PuK & 44\PuK & 0.48 & 91\PuK & 58\PuK &
        11\PuK &
        $176^{+20}_{-23}$ & $138^{+59}_{-43}\PuK$ \nl   
            
Q95 \CAP 3pt & 24 & 101\PuK & 78\PuK & 63\PuK & 1.29 & 49\PuK & 8\PuK &
        11\PuK &
        $64^{+12}_{-27}$ & $65^{+24}_{-14}\PuK$ \nl   
     4pt & 24 & 65\PuK  & 56\PuK & 33\PuK & 1.02 & 32\PuK & 11\PuK  &
        10\PuK &
        $83^{+13}_{-29}$ & $38^{+16}_{-8 }\PuK$ \nl  
     5pt & 24 & 62\PuK  & 43\PuK & 44\PuK & 0.81 & 55\PuK & 6\PuK   &
        9\PuK &
        $108^{+8}_{-32}$ & $54^{+19}_{-10}\PuK$ \nl  
     6pt & 48 & 76\PuK  & 58\PuK & 49\PuK & 0.76 & 64\PuK & 8\PuK   &
        8\PuK &
        $135^{+6}_{-37}$ & $64^{+18}_{-11}\PuK$ \nl   
     7pt & 48 & 74\PuK  & 53\PuK & 52\PuK & 0.67 & 78\PuK & 7\PuK   &
        10\PuK &
        $158^{+7}_{-38}$ & $72^{+19}_{-12}\PuK$ \nl   
     8pt & 48 & 76\PuK  & 58\PuK & 50\PuK & 0.59 & 84\PuK & 9\PuK   &
        10\PuK &
        $178^{+6}_{-38}$ & $81^{+20}_{-12}\PuK$ \nl   
     9pt & 48 & 69\PuK  & 54\PuK & 43\PuK & 0.54 & 79\PuK & 10\PuK  &
        9\PuK &
        $197^{+8}_{-37}$ & $76^{+20}_{-13}\PuK$ \nl   
    10pt & 48 & 70\PuK  & 60\PuK & 37\PuK & 0.50 & 74\PuK & 16\PuK  &
        11\PuK &
        $217^{+6}_{-38}$ & $76^{+22}_{-16}\PuK$ \nl   
    11pt & 48 & 63\PuK  & 56\PuK & 31\PuK & 0.46 & 66\PuK & 18\PuK  &
        11\PuK &
        $237^{+8}_{-37}$ & $83^{+22}_{-15}\PuK$ \nl   
    12pt & 48 & 80\PuK  & 60\PuK & 53\PuK & 0.42 & 126\PuK& 13\PuK  &
        11\PuK &
        $257^{+8}_{-36}$ & $112^{+27}_{-18}\PuK$ \nl   
    13pt & 48 & 78\PuK  & 59\PuK & 51\PuK & 0.40 & 127\PuK& 14\PuK  &
        11\PuK &
        $277^{+10}_{-35}$ & $115^{+29}_{-19}\PuK$ \nl   
    14pt & 48 & 74\PuK  & 64\PuK & 38\PuK & 0.37 & 102\PuK& 24\PuK  &
        11\PuK &
        $297^{+10}_{-34}$ & $74^{+26}_{-24}\PuK$ \nl   
    15pt & 48 & 78\PuK  & 73\PuK & 28\PuK & 0.35 & 79\PuK & 48\PuK  &
        10\PuK &
        $316^{+11}_{-34}$ & $124^{+35}_{-26}\PuK$ \nl   
    16pt & 48 & 96\PuK  & 73\PuK & 63\PuK & 0.33 & 190\PuK& 22\PuK  &
        8\PuK &
        $333^{+13}_{-31}$ & $110^{+38}_{-38}\PuK$ \nl   
    17pt & 48 & 95\PuK  & 78\PuK & 53\PuK & 0.31 & 172\PuK& 30\PuK  &
        8\PuK &
        $357^{+13}_{-31}$ & $110, \ <197$ \nl   
    18pt & 48 & 97\PuK  & 84\PuK & 48\PuK & 0.29 & 166\PuK& 41\PuK  &
        8\PuK &
        $382^{+12}_{-30}$ & $88, \ <191$ \nl   
    19pt & 48 & 99\PuK  & 91\PuK & 40\PuK & 0.27 & 150\PuK& 63\PuK  &
        8\PuK &
        $404^{+13}_{-29}$ & $0, \ <163$ \nl   
   
Q95 \RING 3pt & 96 & 107\PuK & 77\PuK  & 74\PuK & 1.14 & 65\PuK & 5\PuK
        &  7\PuK &
        $170^{+66}_{-54}$ & $59^{+11}_{-9 }\PuK$ \nl  
         4pt & 96 & 104\PuK & 78\PuK  & 68\PuK & 0.85 & 80\PuK & 8\PuK
        &  9\PuK &
        $234^{+67}_{-52}$ & $78^{+16}_{-12}\PuK$ \nl   
         5pt & 96 & 101\PuK & 89\PuK  & 48\PuK & 0.70 & 69\PuK & 18\PuK
        &  8\PuK &
        $286^{+79}_{-39}$ & $69^{+20}_{-21}\PuK$ \nl   
         6pt & 96 & 106\PuK & 107\PuK & 0\PuK &  0.59 & 0\PuK  & \ldots\PuK &
        8\PuK &    
        $332^{+99}_{-16}$ & $19, \ <90$ \nl  

        \tablecomments{ \small Results from each synthesized beam from
          SK93, SK94 and SK95.  The SK93 entries are from Netterfield
          et al. (1995).  The \Ka94 and Q94 entries differ slightly from
          those found in Netterfield (1994) due to the inclusion of East
          to West correlations in the analysis.  $\Delta_{total}$ is the
          raw rms of the binned data.  $\Delta_{inst}$ is the rms
          predicted by the data covariance matrix, $C_D$.
          $\Delta_{sky}^2=\Delta_{total}^2-\Delta_{inst}^2$ gives the
          sky rms.  $\delta T_\ell(\Delta_{sky})$ gives the band power
          inferred from $\Delta_{sky}$ using equation (\ref{eq:rms}).
          $\sigma^{inst}_{\delta T_\ell}$ estimates the contribution to
          the error bars due to the instrument noise and
          $\sigma^{SV}_{\delta T_\ell}$ estimates the contribution to
          the error bar due to sample variance.  The limits on $\ell_e$
          are for where the window function falls to $e^{-1/2}$ of the
          peak value.  The final column lists $\delta T_\ell$ from the
          full likelihood analysis.  The overall $14\%$ calibration
          uncertainty is not included in the errors.} \enddata
 \end{deluxetable}

 The angular spectrum is presented in three ways.  The first is to
 find $\delta T_{\ell_e}$ for the data set associated with each of the 37
 synthesized beams, $H_i$, described in Section~\ref{sec:dTl}.  The
 results of this are presented in Table~\ref{table:dTlsep} and
 Figure~\ref{fig:dTlsep}.  Also included in Table~\ref{table:dTlsep} are
 results from an analysis of $\Delta$ as described in
 Section~\ref{sec:rms}.  Note that for SK95 3pt through 10pt, the
 estimated contribution to the uncertainties due to sample variance,
 $\sigma^{SV}_{\delta T_\ell}$, is approximately equal to the
 contribution due to instrument noise, $\sigma^{inst}_{\delta T_\ell}$.
 Consequently, to significantly reduce the error bars, both better
 sensitivity and more sky coverage will be required.  This
 representation has significant spatial correlations between entries.
 For example, the 3pt beams for Q95, Q94 and \Ka94 all observe the
 same sky and at a similar angular scales.  It is therefore
 inappropriate to combine entries simply by taking the weighted mean
 without taking into account the spatial correlations.

\begin{figure}[tbp]
\plotone{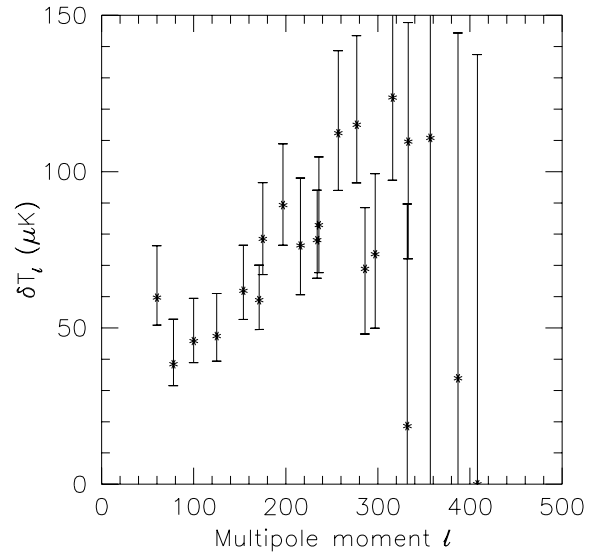}
\caption{The angular spectrum grouped by spatial symmetry.
  These results are given in Table~\ref{table:dTlpt}.}
\label{fig:dTlpt}
\end{figure}

\begin{table}[tbp]
\begin{center}
\begin{tabular}{lcc}  
\hline\hline
        & $\ell_e$ & $\delta T_\ell$ \\
\hline
\B3pt \CAP &     \B60  & $\B59^{+17}_{-9}\uK$ \\  
   \B4pt &      \B78  & $\B38^{+14}_{-7}\PuK$ \\        
   \B5pt &      100 &   $\B46^{+14}_{-7}\PuK$ \\  
   \B6pt &      125 &   $\B47^{+14}_{-8}\PuK$ \\  
   \B7pt &      154 &   $\B62^{+15}_{-9}\PuK$ \\  
\B3pt \RING &    171 &   $\B59^{+11}_{-9}\PuK$ \\  
\B8pt \CAP  &    175 &   $\B79^{+18}_{-11}\PuK$ \\  
   \B9pt &      197 &   $\B89^{+20}_{-13}\PuK$ \\  
    10pt &      216 &   $\B76^{+22}_{-16}\PuK$ \\
\B4pt \RING &    234 &   $\B78^{+16}_{-12}\PuK$ \\
11pt \CAP &      236 &   $\B83^{+22}_{-15}\PuK$ \\
    12pt &      257 &   $112^{+27}_{-18}\PuK$ \\
    13pt &      277 &   $115^{+29}_{-19}\PuK$ \\
    14pt &      297 &   $\B74^{+26}_{-24}\PuK$ \\
\B5pt \RING &    286 &   $\B69^{+20}_{-21}\PuK$ \\
15pt \CAP &      316 &   $124^{+35}_{-26}\PuK$ \\
    16pt &      333 &   $110^{+38}_{-38}\PuK$ \\
    17pt &      357 &   $110, \ <197$ \\
\B6pt \RING &    332 &   $\B19, \ <90$ \\
18pt \CAP &      382 &   $\B88, \ <191$ \\
    19pt &      404 &   $\B\B0, \ <163$ \\
\hline
\end{tabular}
\caption{The angular spectrum grouped by spatial symmetry.  The
  $14\%$ overall calibration uncertainty is not included in the errors.
  \label{table:dTlpt}} 
\end{center}
 \end{table}

 The second way the data are presented is to combine data from spatially
 correlated synthesized beams between years.  To do this, a covariance
 matrix $C_D+C_T$ is made which includes all of the data corresponding
 to a given spatial symmetry.  For example, the \Ka94, Q94 and Q95 3pt
 data forms a $144\times144$ matrix.  The band power is then evaluated
 using a maximum likelihood test.  An effective window function is
 generated using equation (\ref{eq:sumWin}).  The \Ka93 3pt data has largest
 spatial overlap with the 94/95 4pt data and is included there.  These
 results are presented in Table~\ref{table:dTlpt} and
 Figure~\ref{fig:dTlpt}.  The tendency for a rising spectrum is evident
 here though the scatter in the points, particularly at high $\ell$, is
 still considerable.

\begin{figure}[tbp]
  \plotone{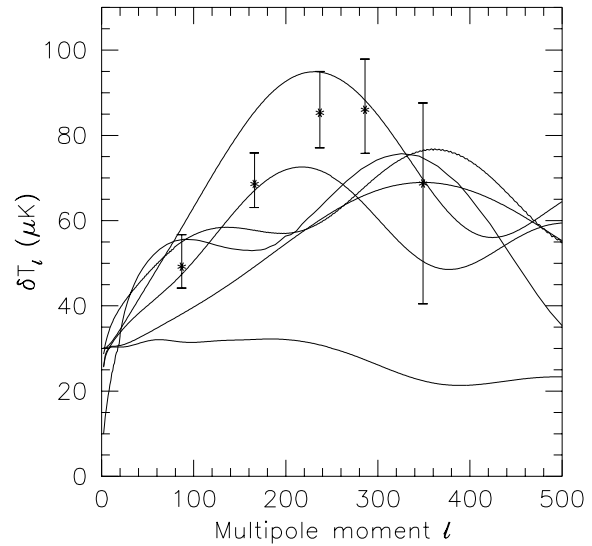}
  \caption{
    The angular spectrum in 5 bins.  The results are tabulated in
    Table~\ref{table:dTlG} and the effective window functions are shown
    in Figure~\ref{fig:winG}.  Predicted spectra from six representative
    theories are also shown.  They are, from top to bottom at $\ell=160$
    a flat $\Lambda$+CDM Model with $\Omega_\Lambda=0.7$ (Ratra and
    Sugiyama 1995), Standard CDM (Ratra et al. 1995), a Texture model
    (Crittenden and Turok, 1995) a PPI model (Peebles, 1995) an
    $\Omega=0.4$ open model (Ratra et al. 1995), and a model with
    reionization (Sugiyama 1995).  The texture model is arbitrarily
    normalized.  All others are normalized to COBE.  To accurately
    compare each of these spectra with the data, they must be convolved
    with the window functions in Figure~\ref{fig:winG}.  The $14\%$
    overall calibration uncertainty is not included in the error bars.
    This effects the normalization of the spectrum, but not the shape.}
  \label{fig:dTlG}
 \end{figure}

\begin{figure}[tbp]
  \plotone{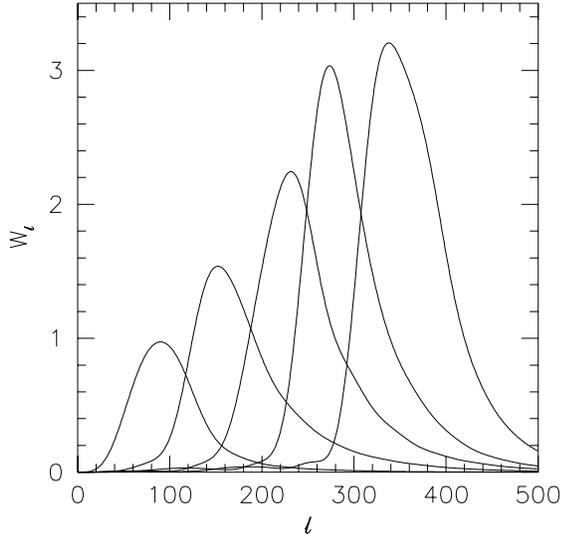}
  \caption{The 5 bin effective window functions.  These have been
    generated from the window functions of the contributing effective
    antenna patterns using equation (\ref{eq:sumWin}).}
  \label{fig:winG}
\end{figure}

\begin{table}[tbp]
\begin{center}
\begin{tabular}{ccccc}  
\hline\hline
$\ell_e$ & $\delta T_\ell$ (\uK) & $\delta T_\ell$ (\uK) & $\delta T_\ell$ (\uK) & $m$\\
         & Low Cut           & High Cut         & Definitive & \\  
\hline
$\B87^{+39}_{-29}$ & $\B47^{+8}_{-5\B}$ & \ldots & $49^{+8}_{-5\B}$ &
        $-0.8^{+1.2}_{-0.1}$ \\
$166^{+30}_{-43}$ & $\B72^{+8}_{-7\B}$  & \ldots      & $69^{+7}_{-6\B}$ &
        $\B0.5^{+0.4}_{-0.4}$   \\
$237^{+29}_{-41}$ & $102^{+11}_{-11}$   & $87^{+9}_{-8\B}$    &
        $85^{+10}_{-8\B}$ & $-0.4^{+0.5}_{-0.2}$        \\
$286^{+24}_{-38}$ & $102^{+12}_{-15}$   & $83^{+10}_{-9}$  &
        $86^{+12}_{-10}$ & $-0.5^{+0.7}_{-0.2}$ \\
$349^{+44}_{-41}$ & $\B93^{+23}_{-32}$  & $60^{+15}_{-22}$ &
$69^{+19}_{-28}$ & \ldots       \\
\hline
\end{tabular}

\caption{The angular spectrum in 5 bins. The first bin is
  comprised of the 3pt to 6pt \CAP data from all three years.  The
  second is from the 7pt to 9pt \CAP data and the 3pt \RING
  data.  The third is 10pt to 12pt \CAP and 4pt \RING.  The
  fourth is 13pt to 15pt \CAP and 5pt \RING.  The fifth is
  16pt to 19pt \CAP and 6pt \RING.  See
  Section~\ref{sec:data} for a description of the cut levels.  $m$ is
  the angular spectral index within each bin. \label{table:dTlG}}
\end{center}
 \end{table}

 Finally, the data are combined into 5 angular scale bins and presented
 in Table~\ref{table:dTlG} and Figure~\ref{fig:dTlG}.  To produce each
 of these bins, all of the data which is sensitive to a given range of
 angular scales is combined using the full correlation matrix (see
 Section~\ref{sec:like}).  For instance, the first bin is comprised of
 the 3pt to 6pt \CAP data from all three years.  The effective window
 functions for each bin using equation (\ref{eq:sumWin}) are presented in
 Figure~\ref{fig:winG}.

 Also presented in Table~\ref{table:dTlG} are values for the angular
 spectral index $m$ (see Section~\ref{sec:dTl}).  The negative spectral
 index around $\ell\approx80$ which was reported in Netterfield (1994)
 with the 1994 data alone is again seen in the lowest angular bin.
 The lowest four points in Figure~\ref{fig:dTlpt} also shows this
 locally falling spectrum.  However, this trend is dominated by the 3pt
 data being high.  Additionally, for each of the 5 angular bins,
 including the first one, the slope is consistent with 0.  The angular
 spectrum of the whole sky may have a dip at these
 scales, or this effect could be a result of sample variance.  This
 issue will be resolved by extended sky coverage to reduce sample
 variance.

 While all known correlations have been taken into account in generating
 the likelihoods for each of the combined groups, there are still some
 correlations between groups.  They are small ($<0.2$), however, and may
 be ignored for most analysis.

 Not included in the quoted error bars are the effects of calibration
 and beam uncertainties.  As discussed, the calibration uncertainty is
 $14\%$ ($1\sigma$), mainly due to uncertainty in the temperature scale
 of Cas-A.  This has very little effect on the uncertainty of the
 spectral index, $\beta$, and no effect on the shape of the angular
 spectrum.  Rather this has only an effect on the over-all
 temperature scale.  This large normalization uncertainty must be
 included in any comparison with theoretical predictions.

 Beam uncertainties, as well as contributing to calibration errors, also
 produce uncertainties in the angular spectrum.  For uncertainties
 in the beamwidth along the elevation axis, the only effect is in the
 calibration (miss-estimating the beam solid angle in predicting the
 signal from Cas-A), and has been included in that error bar.  For
 uncertainties in the azimuth beam width the effect is a
 function of the weighting vectors used.  For the 4pt synthesis, where
 the physical beam width is small compared to the width of the
 synthesized lobes, only the temperature scale is effected, as with the
 elevation beam.  For the 13pt synthesis, however, the additional effect
 of mis-stating the beam overlap between points very nearly cancels the
 effect on the calibration.  The result of this is that the $2\%$
 azimuth beam uncertainty yields an additional $2\%$ relative
 uncertainty between the 4pt and 13pt synthesized beam data.  This
 effect is small for this experiment given the error bars on the
 spectrum and is ignored.\\ 

\section{Foreground Contamination} \label{sec:for}

\begin{deluxetable}{lccccc}
\tablecaption{Bright Point Sources\label{table:sources}}
\tablehead{
\colhead{Source} & \colhead{$\nu_o$}  & \colhead{Flux at $\nu_o$}
 & \colhead{Spectral Index $\alpha$} & \colhead{Flux at 40\,GHz} &
\colhead{Reference}}
\startdata
0014+81 & 10.7\,GHz & 0.73 Jy & 0.36 &  1.2\,Jy & S5 \nl
0210+860 (3C61.1) & 30.0 & 1.0 & 0.2 &  1.1     & HR \nl
0454+844 & 10.0 & 1.5 & 0.0 & 1.5               & K \nl
0615+820 & 10.7 & 0.86 & 0.0 &  0.86    & E,K\nl
0740+82  & 10.7 & 0.65 & -0.47 & 0.35   & HR, S5 \nl
1003+83  & 10.7 & 0.66 & -0.10 & 0.58   & S5 \nl
1039+811 & 90.0 & 0.80 & 0.0 & 0.80     & IRAMc \nl
1053+815 & 20.0 & 0.4  & 0.43 & 0.54    & E,VLBIc \nl
1050+812 & 10.7 & 1.1 & -0.1 & 0.96     & K \nl
1221+809 & 10.7 & 0.6 & +0.2 & 0.78     & S5, VLAc \nl
1637+8239 (NGC6251) & 10.7 & 0.8 & -0.4 & 0.47  & HR \nl
1637+826 & 10.7 & 0.73 & -0.4 & 0.43    & S5,VLAc \nl
2342+821 & 30.0 & 0.8 & 0.2 & 0.85      & HR, S5 \nl
\enddata

\tablecomments{ Listed are parameters for the brightest point sources
  used in the estimate of the level of point source contamination.  The
  flux at 40\,GHz is extrapolated from the measurement at $\nu_o$ by
  $S(\nu)=S(\nu_o)(\nu/\nu_o)^\alpha$, where $\alpha=\beta+2$.  The
  measured value from the literature for $\alpha$ used where possible.
  Otherwise $\alpha=0$ (flat spectrum or constant flux with frequency)
  is assumed.  The data are obtained from \cite{Eck86} (E), \cite{Her92}
  (HR), \cite{Ste92} (IRAMc), \cite{Kur81a} (K), \cite{Kur81b} (S5),
  \cite{Per82} (VLAc) and \cite{Ede87} (E).  }

\end{deluxetable}

\begin{figure}[tbp]
\plotone{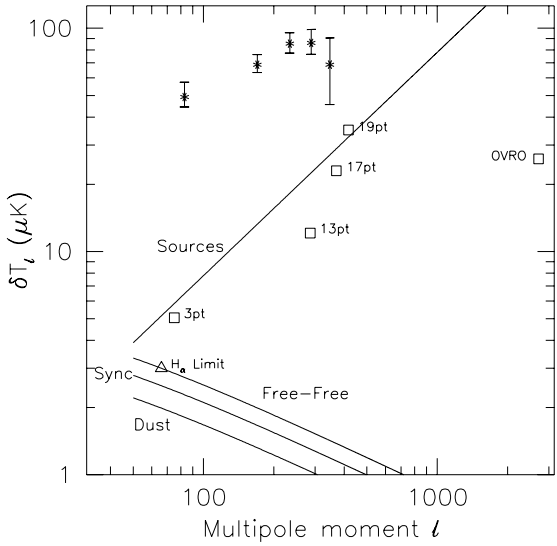}
\caption{Estimated levels of foreground contamination at 40\,GHz.  The 5
  bin data from this experiment are given as asterisks.  The estimated
  angular spectra for the various foreground sources are from Tegmark
  and Efstathiou (1995).  The free-free angular spectrum has been
  renormalized for our observing region using a limit based on
  H$_\alpha$ emission (Gaustad et al. 1995; Simonetti et al. 1995).  An
  upper limit for the normalization of the synchrotron spectrum has been
  estimated by extrapolating the 408\,MHz Haslam map to 40\,GHz for our
  observing region (Haslam et al. 1982, Wollack et al. 1993).  The
  square boxes represent contributions to our data estimated by
  extrapolating exiting point source surveys to 40\,GHz (see text).
  Also indicated is an upper limit on sources at smaller scales based on
  the OVRO Ring experiment (Meyers, Readhead, and Lawrence, 1993)
  extrapolated from 20\,GHz.  }
\label{fig:sources}
 \end{figure}

 The angular spectra presented in the previous section are for the
 microwave sky at ~40\,GHz.  There are, however, potential sources of
 foreground contamination.  Two major classes are diffuse Galactic
 emission, and unresolved point sources.  Limits may be placed on the
 level of both of these potential contaminants.

 Potential sources of diffuse Galactic contamination include dust,
 synchrotron, and free-free emission.  The level of dust contamination
 at 40\,GHz and $\ell=100$ has be estimated to be less than 2\uK
 (\cite{Teg95}), which is far smaller than the signals measured here.
 Similarly, recent measurements of the amplitude of diffuse free-free
 emission based on H$_\alpha$ emission have been made for the Saskatoon
 observing region (\cite{gau95}; \cite{Sim95}).  These measurements
 place an upper limit of 6\uK at 30\,GHz for the SK93 3pt beam.  Once
 again this is too small to account for the observed signals.  An
 analysis of high latitude Galactic emission in the COBE 2 year data
 places similar limits on both foregrounds at large angular scales
 (\cite{Kog95}).  An estimate of the amplitude of synchrotron emission
 can be made by extrapolating radio maps at 408\,MHz (\cite{Has82}) and
 1.4\,GHz (\cite{Rei86}) to 30\,GHz assuming a temperature spectral
 index of $\beta=-2.8$.  This predicts a signal of less than 5\uK in
 \Ka93, which is much smaller than the measured signals.  However, if
 the spectral index of the synchrotron emission varies spatially by
 $\delta\beta\approx0.05$, then $30\uK$ signals could be expected.  The
 spectral index of these fluctuations would still be characterized by
 $\beta\approx-2.8$.

 The most powerful discriminant against foreground contamination is
 based on the frequency spectrum of the fluctuations ($\beta$).  In
 Section~\ref{sec:beta}, for $\ell\approx73$, $\beta$ was
 found to be $0.2\pm0.3$ which is inconsistent with the spectral index
 of dust ($\beta=1.7$), free-free emission ($\beta=-2.1$), or
 synchrotron emission ($\beta=-2.8$).  Additionally, since the angular
 spectrum for galactic emission falls as $\delta T_\ell \propto\ell^{-1/2}$
 (\cite{Teg95}), it is very unlikely that the galaxy is a significant
 contaminant at any angular scale.

 Of more concern for several reasons is potential contamination by
 unresolved point sources.  Firstly, the angular spectrum of a
 family of point sources is expected to rise as $\delta T_\ell \propto
 \ell$, so the $\beta$ limits placed at large angular scales are weaker
 at smaller angular scales.  Additionally, since the mechanism for
 radiation varies between individual sources, the expected $\beta$ also
 varies, and in many cases is not accurately described by a simple power
 law.  This is of particular concern since no flux-limited survey of the
 SK observing region in has been performed over our frequencies (26\,GHz
 to 40\,GHz).  However, it is still possible to make some estimates of
 the level of contamination via comparison with \MS, source counts
 from the OVRO Ring experiment (\cite{Mey93}) at smaller scales and an
 extrapolation of existing point source surveys.

 While $\beta$ varies from source to source, it is expected that over a
 sufficiently wide frequency range no foreground source will have a
 spectrum consistent with $\beta=0$.  Consequently our comparison of
 SK95 with \MS\ which yields a limit on $\beta$ of $0.1\pm0.2$ with a
 frequency baseline from 40\,GHz to 240\,GHz assures that the sky
 covered by both experiments is devoid of significant point
 contamination.  However, this field only represents 8 square degrees of
 the total of 200 square degrees observed by SK95.

 A second approach is to use results from other experiments and point
 source surveys.  The OVRO Ring experiment has observed 96 fields at
 20\,GHz.  They find $\delta T_\ell=103\uK$ at $\ell_e=2017$.  Here we
 make the assumption that the signal is dominated by Poisson
 distributed unresolved point sources.  Extrapolating in angular scale
 by $\delta T_\ell\propto\ell$ to $\ell=404$, the angular scale of the
 19pt data set, one arrives at an expected contribution of 21\uK.  Since
 OVRO Ring was performed at 20\,GHz, this number needs to also be scaled
 in frequency.  A conservative estimate of the `typical' $\beta$ for a
 point source is that its flux is constant in frequency, (i.e.,flat
 spectrum or $\beta=-2$).  Thus, an estimate of the point source
 contribution at the smallest scales of the SK experiment based on OVRO
 Ring source counts is 10\uK at 40\,GHz.

 As a final estimate of the level of point source contamination, the
 expected amplitude of known point sources in the observing region is
 extrapolated.  Table~\ref{table:sources} lists the brightest of these.
 In addition to the sources listed in Table~\ref{table:sources}, all of
 the sources in the S5 catalog (\cite{Kur81b}) are used.  Based on an
 analysis of IRAS faint sources, it is considered unlikely that a new
 population of sources exists within our observing frequencies
 (\cite{Con95}).  The spectral index from the literature is used to
 extrapolate to 40\,GHz whenever possible.  If a measured frequency
 spectral index is not known, the source is assumed to be of constant
 flux with frequency.  A source map of the observing area is made, and
 the SK observing strategy simulated on it.  The largest estimated
 contribution is to the 19pt data, and is $\delta T_\ell=35\uK$, which
 could be significant.  However, from Table~\ref{table:dTlsep}, the 19pt
 data yields a 163\uK upper limit with a maximum likelihood at 0\uK.
 The estimated contribution to the 17pt data is 23\uK, which would imply
 a 4\% contribution to our 69\uK detection at these scales.  The
 predicted signal is dominated by sources in RA bins 5 and 10.  The
 measured 17pt data show no excess in these bins, and re-analyzing the
 data with these bins deleted changes the most likely value for the 17pt
 data by less than 1\%.  For the 13pt data, at $\ell=277$, the expected
 contribution is $\delta T_\ell=12\uK$.  From Table~\ref{table:dTlsep},
 the measured value is $115\uK$.  Subtracting the source contribution in
 quadrature yields less than a 1\% effect.

 The data, along with estimated contributions from foreground
 contaminants are presented in Figure~\ref{fig:sources}.  The
 indications are that the SK data are not seriously contaminated by
 foreground sources.  However, extending the coverage to smaller angular
 scales will require improved discrimination from point sources.
 
 \section{Discussion}

 The rich set of internal and external consistency checks inherent to
 this experiment indicate that a signal fixed to the celestial sphere
 has been detected. The limits placed on the frequency spectral index
 $\beta$ rule out significant contamination from diffuse Galactic
 emission.  The limits on $\beta$ over a very large frequency range
 placed by the successful comparison with the \MS\ experiment further
 reduce the likelihood of significant contribution to the signal by
 foreground contaminants.  Conservative extrapolations from existing
 point source catalogs strengthen this argument.  The evidence is
 compelling that the signal which has been detected is due to
 anisotropies in the CMB.

 This experiment has measured the angular spectrum of these
 anisotropies simultaneously with one experiment and with equal
 calibration error at all angular scales.  We find that the spectrum
 rises from $\delta T_\ell = 49^{+8}_{-5}$\uK at $\ell = 87$ to
 $85^{+10}_{-8}\uK$ at $\ell = 237$.  This can be used to place
 significant limits on the reionization history of the universe, and,
 within a given theoretical framework, begin to place limits on
 cosmological parameters.  A quantitative comparison with theory will
 presented elsewhere.

 Given the relative complexity of this analysis, it is important to
 mention that much of the analysis has been repeated with completely
 different programs by two of the authors to verify accuracy.  The
 consistency of the naive $\Delta$ based tests with the full likelihood
 analysis further strengthens our confidence in our programs and
 techniques.  In addition, Andrew Jaffe and J.R. Bond at CITA have
 repeated and extended our analysis of the SK93 and SK94 data, again
 confirming our analysis.

 To improve the discriminating power of this experiment several things
 are needed besides reducing the size of the error bars.  First of all,
 it must be confirmed at high $\ell$.  Additionally, a wider frequency
 coverage at smaller scales will improve discrimination against point
 sources.  A better calibration will improve comparison with COBE,
 effectively increasing the range of angular scales over which a theory
 can be tested.  It is possible to make a map of the CMB anisotropy from
 these data.  A map of an extended region of the sky with better signal
 to noise will be useful to distinguish between Gaussian and
 non-Gaussian models.

 We are indebted to Dave Wilkinson for his work on the experiment, and
 for many very useful discussions.  We thank Marion Pospieszalski and
 Mike Balister at NRAO for providing the HEMT amplifiers upon which this
 experiment was based.  George Sofko and Mike McKibben at the University
 of Saskatchewan and Larry Snodgrass at the Saskatchewan Research
 Council provided valuable site support in Saskatoon.  This work has
 benefited from useful discussions with Dick Bond, Ruth Daly, Tom
 Herbig, Andrew Jaffe, Jim Peebles, Bharat Ratra, Dave Spergel, and Paul
 Steinhardt.  Financial support has been provided by NSF grant PH
 89-21378, NASA Grants NAGW-2801 and NAGW-1482, a Research Corporation
 Award, a Packard Fellowship, and an NSF NYI grant to L. Page.

 The data, beam profiles, window functions and covariance matrices will
 be made publicly available upon acceptance of this paper.

\end{document}